\documentclass[acmtog]{acmart}
\renewcommand\footnotetextcopyrightpermission[1]{} % removes footnote with conference information in first column

\usepackage{amsmath}
\usepackage{booktabs}

\newcommand{\dd}{\mathrm{d}}

\newcommand{\bomega}{\bm{\omega}}
\newcommand{\bs}{\mathbf{s}}

\newcommand{\bd}{\mathbf{d}}

\def\figurePath{}
\def\myfigure#1#2#3{\begin{figure}[htb]\centering\includegraphics[width = \linewidth]{\figurePath#2}\caption{#3}\label{fig:#1}\end{figure}}
\def\mycfigure#1#2#3{\begin{figure*}[htb]\centering\includegraphics*[clip, width = \linewidth]{\figurePath#2}\caption{#3}\label{fig:#1}\end{figure*}}

\usepackage{bm}

\makeatletter
\newcommand\paragraphNew{\@startsection{paragraph}{4}{\parindent}%
  {-.5\baselineskip \@plus -2\p@ \@minus -.2\p@}%
  {-3.5\p@}%
  {\ACM@NRadjust{\@parfont}}}
	\makeatother
	
\AtBeginDocument{%
  \providecommand\BibTeX{{%
    \normalfont B\kern-0.5em{\scshape i\kern-0.25em b}\kern-0.8em\TeX}}}

\begin{document}

% \title{Position-Free Path Tracing for Multiple Bounce Scattering}
% \title{Full-spherical, Position-free and Multiple-bounce Formulation of Smith Microfacet BSDFs}
% \title{Position-free Multiple-bounce Microfacet BSDFs}
\title{Position-free Multiple-bounce Computations for Smith Microfacet BSDFs}

\author{Beibei Wang}
\affiliation{School of Computer Science and Engineering, Nanjing University of Science and Technology}
\author{Wenhua Jin}
\affiliation{School of Computer Science and Engineering, Nanjing University of Science and Technology}
\author{Jiahui Fan}
\affiliation{School of Computer Science and Engineering, Nanjing University of Science and Technology}
\author{Jian Yang}
\affiliation{School of Computer Science and Engineering, Nanjing University of Science and Technology}
\author{Nicolas Holzschuch}
\affiliation{INRIA}
\author{Ling-Qi Yan}
\affiliation{University of California, Santa Barbara}

\renewcommand{\shortauthors}{Wang, et al.}

\begin{abstract}
Bidirectional Scattering Distribution Functions (BSDFs) encode how a material reflects or transmits the incoming light. The most commonly used model is the Microfacet BSDF. It computes material response from the micro-geometry of the surface assuming a single bounce on specular microfacets. The original model ignores multiple bounces on the micro-geometry, resulting in energy loss, especially with large roughness. In this paper, we present a position-free formulation of multiple bounces inside the micro-geometry, which eliminates this energy loss. We use an explicit mathematical definition of path space that describes single and multiple bounces in a uniform way. We then study the behavior of light on the different vertices and segments in path space, leading to an accurate and reciprocal multiple-bounce description of BSDFs. We also present practical, unbiased Monte-Carlo estimators to compute multiple scattering. Our method is less noisy than existing algorithms for computing multiple scattering. It is almost noise-free with a very-low sampling rate, from 2 to 4 samples per pixel. 
\end{abstract}

\begin{CCSXML}
<ccs2012>
	 <concept>
				<concept_id>10010147.10010371.10010372</concept_id>
				<concept_desc>Computing methodologies~Rendering</concept_desc>
				<concept_significance>500</concept_significance>
	 </concept>
   <concept>
       <concept_id>10010147.10010371.10010372.10010376</concept_id>
       <concept_desc>Computing methodologies~Reflectance modeling</concept_desc>
       <concept_significance>500</concept_significance>
       </concept>
 </ccs2012>
\end{CCSXML}

\ccsdesc[500]{Computing methodologies~Rendering}
\ccsdesc[500]{Computing methodologies~Reflectance modeling}
\keywords{microfacet, position-free, multiple-bounce, full-spherical}

\begin{teaserfigure}
  \centering
  \includegraphics[width=\textwidth]{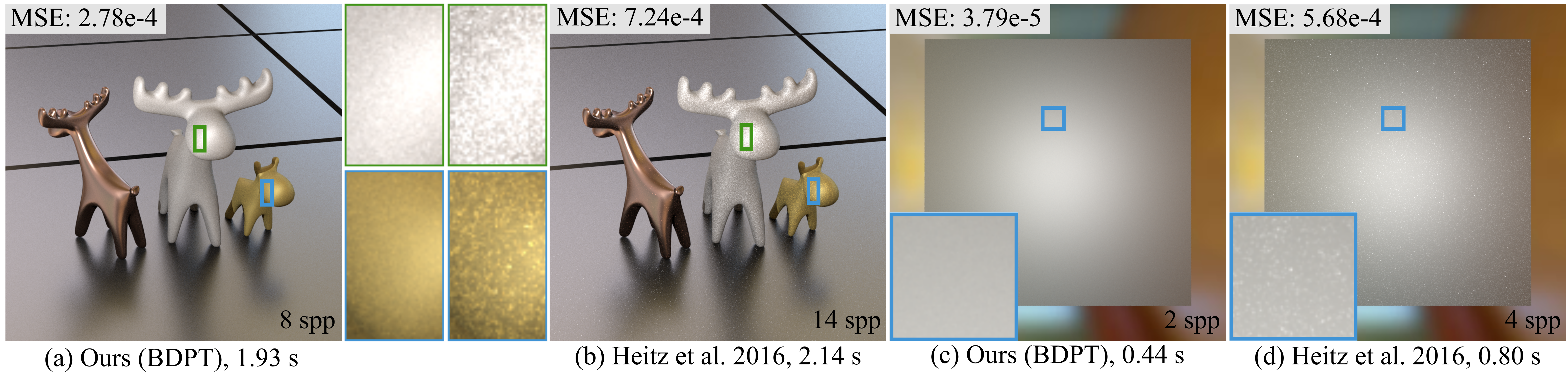}
  \caption{We present a new, position-free, method to compute multiple bounces inside the micro-geometry for the micro-facet BSDF model. Our method produces less noise than previous work (Heitz et al.~\shortcite{heitz2016}), for both rough conductors and rough dielectrics. (a) and (b): equal-time comparison with Heitz et al.~\shortcite{heitz2016} for rough conductors (from left to right: copper (GGX model, $\alpha$ = 0.1), aluminum (GGX model, $\alpha$ = 0.6) and gold (GGX model, $\alpha$ = 0.5)). (c) and (d): comparison with Heitz et al.~\shortcite{heitz2016} for a rough dielectric (GGX model, $\alpha$ = 1.0). In both cases, our method reduces noise significantly (see insets).}
  \label{teaser}
\end{teaserfigure}

\maketitle

\section{Introduction}
\label{sec:intro}

Material properties or reflectance encode how materials interact with the incoming light. Having good material properties is essential for photorealistic rendering. Microfacet models are widely used both in real-time applications and in high-quality offline rendering. They predict the appearance of the material from the statistical properties of the micro-geometry of its surface. The most common model, Cook-Torrance~\cite{Cook1982REFLEC, walter2007mmrt}, assumes that the surface is made of planar specular microfacets, and computes the material response by integrating a single bounce over this micro-geometry. The BRDF is connected to the distribution of microfacet normals. By nature, the Cook-Torrance model ignores the light that has bounced several times on the micro-geometry, resulting in an energy loss. The effect is especially visible when the surface has a high roughness. 

Multiple algorithms have tried to enhance the microfacet models by computing multiple bounces. The common idea is that the shadowing-masking term of the Cook-Torrance model actually encodes the proportion of light that was not reflected in the first bounce. This light that was blocked by the micro-geometry is used as input to compute the multiple-bounce term. 

There are several models for the shadowing-masking term of the Cook-Torrance BRDF model: the V-groove model assumes that for each microfacet, there is another one next to it forming a V-shaped groove with it; the Smith model assumes the independent distribution of microfacets heights and normals, treating the microfacets as randomly distributed microflakes. The former is easier for numerical analysis and provides explicit analytical solutions, the latter requires a double indefinite integral to compute the shadowing-masking term; some microfacet distributions still have an analytical term for shadowing-masking.

Several algorithms use the V-groove model to compute multiple bounces in the micro-geometry, and provide an analytical formula for the missing energy~\cite{lEE2018PRATMUL,Feng2018multiV}. However, the V-groove model has several issues that can make it undesirable: discontinuities or singularities in the shadowing-masking term, an overall shiny appearance even for very rough surfaces, and the inability to model transparent materials. 

The Smith model results in a better overall appearance for the material, but there is no explicit formula to compute multiple bounces of light. Heitz et al.~\shortcite{heitz2016} showed that it was nevertheless possible to compute multiple scattering effects inside the micro-geometry using random walks. Their simulation takes into account the position of the shading point inside the micro-geometry, leading to a very accurate result, at the expense of computation time. 

In this paper, we present a new understanding of multiple-bounce microfacet BSDFs. Inspired by the position-free approach that Guo et al.~\shortcite{Guo:2018:Layered} applied to layered materials, we analyze and formulate the \emph{path space} as the light undergoes an arbitrary number of bounces inside general microfacet BSDFs. In this path space, we study the behavior of light at the vertices and segments along different paths, introducing the \emph{vertex term} and the \emph{segment term}, respectively. The vertex term and segment term together bring a clean physically-based separation of distribution and occlusion -- the vertex term accounts for the normal distributions and Fresnel effects, while the segment term focuses on shadowing-masking effects and thus leads to energy conservation.

With our explicit position-free formulation, we propose practical Monte Carlo estimators, exploiting path tracing (PT) and bidirectional path tracing (BDPT) to efficiently solve the integration. Our method provides result similar to Heitz et al.~\shortcite{heitz2016}, with a significantly decreased noise level for evaluation-only tasks, even with very low sample counts, as low as $2-4$ spp. It passes the white furnace test. It works with dielectrics, anisotropic materials and commonly-used normal distributions such as Beckmann and GGX. 

\begin{table}[!t]
	\renewcommand{\arraystretch}{1.1}
	\caption{\label{tab:notations} Notations.}
\begin{small}
 % \left
  \begin{tabular}{|l|c|l|}\hline
  \multicolumn{2}{|c|}{Mathematical notation}\\
	\hline
		$\Omega^\pm$      			                                   &full spherical domain \\ 
		$\Omega^+$      			                                     &upper spherical domain \\ 
		$\omega_i \cdot \omega_o$      			                       &dot product \\ 
		$|\omega_i\cdot \omega_o|$      		                     	&absolute value of the dot product \\ 
		$\left\langle\omega_i, \omega_o \right\rangle$      	&dot product clamped to 0 \\ 
		$\chi^{+}(a)$                                                 &Heaviside function: 1 if a > 0 and 0 if a $\leq$ 0 \\
		\hline
		 \multicolumn{2}{|c|}{Physical quantities used in microfacet models}\\
				\hline
		$\omega_g = (0,0,1)$ & geometric normal \\
		$\omega_m $ & microfacet normal \\
		$\omega_i $ & incident direction ($\omega_i \cdot \omega_g$ could be < 0) \\
		$\omega_o $ & outgoing direction \\
		$\Lambda(\omega)  $ &the Smith Lambda function \\
		$D(\omega_m)$ &normal distribution function \\
		$F(\omega_i,\omega_m)$ &Fresnel factor \\		
		$G_1(\omega,\omega_m)$ &masking function \\
		$G_1^{\mathrm{local}}(\omega,\omega_m)$ & masking (local)\\
		$G_1^{\mathrm{dist}}(\omega)$& masking (distant)\\
		$\rho(\omega_i,\omega_o)$& multiple-bounce BSDF\\
		\hline
		\end{tabular}

\end{small}
\end{table}
%-----------------------------------%
\section{Related Work}
\label{sec:related}
%-----------------------------------%
%

\paragraph{Microfacet models.}
Torrance and Sparrow~\shortcite{TorranceTorrance:1967} introduced the microfacet model for reflection on rough surfaces. They extract the overall material reflectance from a statistical description of the surface micro-geometry, made of specular microfacets. The model focuses on a single bounce over this micro-geometry, and gives a full BRDF model from the surface characteristics. The most important parameter is the probability distribution of the microfacet normals (Normal Distribution Function, or NDF). The model depends on two other terms: the Fresnel term, connected to the composition of the material, and the shadowing-masking term, encoding how much of the incoming light goes into this first bounce. The model was introduced to the graphics community by Cook and Torrance~\shortcite{Cook1982REFLEC} and extended to rough dielectrics by Walter et al.~\shortcite{walter2007mmrt}. 

The normal distribution function has a strong impact on the overall aspect of the BRDF. Initial works used the Beckmann distribution~\cite{BeckmannSpizzichino:1963, TorranceTorrance:1967, Cook1982REFLEC}. Trowbridge and Reitz~\shortcite{Trowbridge:75} introduced a different distribution, corresponding to microfacets distributed on half-ellipsoids. It was rediscovered by Walter et al.~\shortcite{walter2007mmrt} as the GGX distribution. Other statistical distributions have been introduced, see e.g. \cite{bagher:2012, Ribardiere:2017:STD}. 

Ashikhmin and Premo\v{z}e~\shortcite{Ashikhmin:2007} extracted the NDF from measured BRDFs. The technique was extended by Dupuy et al.~\shortcite{dupuy:2015} and Ribardière et al.~\shortcite{Ribardiere:2019:NDF}. None of them can solve the multiple-bounce issue of the microfacet model.

\paragraph{Shadowing-masking.} 
The shadowing-masking term is important for energy conservation in the microfacet model. It encodes how much of the incoming light was blocked by the micro-geometry (shadowing) as well as how much of the reflected light was blocked (masking). To compute it, we need a model of the surface micro-geometry. Initial work~\cite{TorranceTorrance:1967,Cook1982REFLEC} relied on the V-groove model: for each microfacet, there is another microfacet facing it with the same slope. The V-groove model results in simple computations, as occlusion only depends on the current microfacet slope. The resulting shadowing-masking term has discontinuous derivatives.

Smith~\shortcite{smith:1967:smith} computes the shadowing-masking term from the NDF, assuming that the orientations and positions of the microfacets are independent. The shadowing-masking term is computed from the NDF through a double integration. The resulting term is smooth, and varies more consistently with the roughness of the NDF. Walter et al.~\shortcite{walter2007mmrt} and Heitz~\shortcite{HEITZ2014UNDER} explain and expand the Smith shadowing-masking term for more distributions and take into account the correlation between incoming and outgoing direction. 

\paragraph{Multiple-bounces in microfacet models.} 

By nature, the microfacet models only express the light reflected after a single bounce on the surface micro-geometry. Light that bounces several times is not represented, resulting in an energy loss. The effect is particularly visible for rough surfaces. Kelemen and Szirmay-Kalos~\shortcite{Kelemen01amicrofacet} introduce multiple-scattering to the microfacet BSDF by computing the portion of light blocked by the shadowing-masking term and reintroducing it as a diffuse component. The method was extended by Jakob et al.~\shortcite{jakob:2014:layered} on dielectric and conductor in layered materials.

Heitz et al.~\shortcite{heitz2016} proposed a multiple-bounce method treating the microfacets randomly distributed microflakes, resulting a random walk solution, which reaches an agreement with the simulated data from surfaces~\cite{heitz2015implement}. Dupuy et al.~\shortcite{Dupuy:2016:Unification} introduced a unified model between multiple bounce in microsurfaces and microflakes. Sch\"{u}ssler et al.~\shortcite{Schussler:2017:normal} extended the approach to normal-mapped surfaces. Westin et al.~\shortcite{Westin:92:RandomWalk} encoded multiple scattering by using random walks in microgeometry, and Falster et al.~\shortcite{Falster:2020:Wave} combined Westin’s approach with wave optics. These methods match the simulated data very well, but do not have an explicit solution. The random walk simulation results in large variance in the rendered results. Compared to theirs, our method has an explicit formula, although our method still relies on Monte Carlo methods to solve this formula. However, without the need of tracing the height during random walks, our method produces less noise. This explicit formulation enables the use of more advanced light transport methods such as Bi-Directional Path Tracing, further reducing the variance.

All these methods use the Smith shadowing model. By contrast, using the V-groove model allows for analytic solutions for multiple-bounce~\cite{lEE2018PRATMUL, Feng2018multiV}. The drawbacks are those of the V-groove model: too shiny for rough surfaces, discontinuous derivatives, and not compatible with transparent materials. Lee et al.~\shortcite{lEE2018PRATMUL} redistribute energy to mask the discontinuities, but at the cost of re-introducing randomness.

Kulla and Conty~\shortcite{KullaConty:2017:revisiting} approximate proposed multiple bounces in microfacets by mixing the single scattering an azimuthally invariant lobe. Turquin~\shortcite{Turquin:2019:multiple} proposed an even simpler multiple bounce computation approach, by scaling the single bounce results. The scaling factor is precomputed based on the surface roughness, the outgoing angle and the index of refraction for dielectrics. These methods are fast, but the multiple bounce term does not have the properties observed in simulations. 

Meneveaux et al.~\shortcite{Meneveaux:2018} proposed an analytical model for the multiple reflections of light between the interface and the substrate for interfaced Lambertian materials, but ignores the multiple reflections between microfacts. Xie et al.~\shortcite{Xie:2019:multiple} proposed to represent the multiple scattering with Gaussians or the Real NVP neural network, and used these models for rendering at run-time. Both of two models produce close to energy-conserving results, but with no performance reported.

\paragraph{Position-free formulation for layered materials.}
The position-free path integral formula was proposed by Guo et al.~\shortcite{Guo:2018:Layered} for the evaluation and sampling of layered materials, and is recently improved by Xia et al.~\shortcite{Xia:2020:Layered} and Gamboa et al.~\shortcite{Gamboa:2020:EfficientLayered} with a more advanced sampling method or a more efficient estimator.

The biggest advantage of the position-free formulation is that, it allows explicit representation of light transport in a subspace. Then, advanced methods and estimators can be exploited to reduce variance. Inspired by this line of work, we formulate the multiple bounce of light transport within BSDFs using the position-free path integral.

%-----------------------------------%
\section{Position-free Multiple-bounce BSDF Formulation}
\label{sec:posfree}
%-----------------------------------%

In this section, we describe our path formulation of general light transport for any bounces of microfacet BSDFs. We first introduce our position-free formulation with the definition of a path, then dive into its components on vertices (vertex term) and segments (segment term). We show that the vertex term controls the local light transport that reflects / refracts according to the Fresnel and NDF, while the segment term is responsible for global light transport that accounts for occlusions and multiple scattering. After that, we present detailed derivations of both termsa and analyze their properties.

\subsection{Motivation}

\myfigure{localSmith}{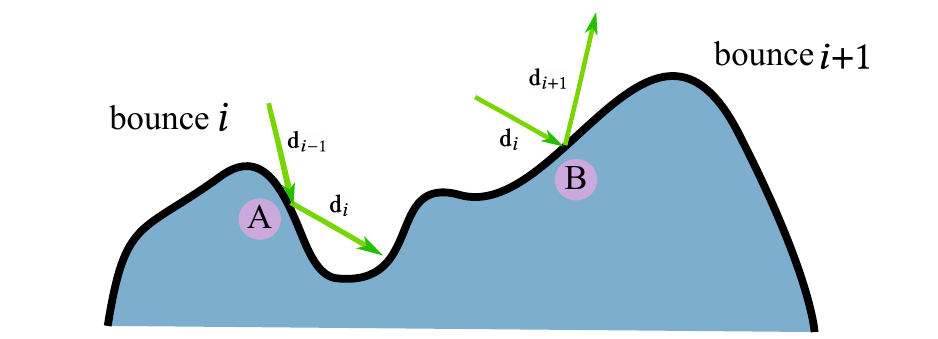}{Multiple-bounce BSDF under our position-free observation. In a local neighborhood, although A and B have different positions and heights, they share the same statistics, especially the NDF, which essentially makes A and B identical. Therefore, bounce $i$ and bounce $i+1$ are processed independently, except that bounce $i+1$ gets the incident direction $\bd_{i}$ from bounce $i$'s outgoing direction $\bd_{i}$. The independence also explains why both directions require the shadowing-masking function -- shadowing from $\bd_i$ to determine at A the proportion of energy able to proceed to the next bounce, and masking from $-\bd_i$ to determine the visible normals at B.}

Our key idea is to use a position-free formulation for the integration of multiple-bounces inside the micro-geometry. It is based on two observations: first, at the macro scale, BSDFs use a position-free formulation: the point at which the incident light arrives and the point from which the outgoing light leaves are considered to be identical, regardless of how many bounces the light undergoes in the micro-geometry. Second, at the micro scale, the Smith shadowing theory~\shortcite{smith:1967:smith} assumes that the positions and normals of the microfacets are uncorrelated. This leads to an obvious but important observation: all points in the micro scale can be considered the same, statistically. BSDFs are also position-free in the micro scale (Fig.~\ref{fig:localSmith}).

Note specifically that the position-free formulation is irrelevant to specific height distributions. Different height distributions (often referred to as $P_1$ in related literature, and typically assumed to be either Gaussian or uniform), are only used to derive the shadowing-masking functions. Interestingly, as discussed in Heitz et al.~\shortcite{heitz2016} and implicitly suggested in the Appendix of Walter et al.~\shortcite{walter2007mmrt}, the choice of height distribution functions does not even affect the final result of the shadowing-masking functions, since the height distribution is canceled out in the derivation. Our method, following the position-free formulation, is independent of any specific type (e.g. uniform) of height distributions, similar to previous work.

A direct consequence of the position-free formulation is that the outgoing direction for the current bounce is identical to the incoming direction for the next bounce. This leads to two considerations: First, the occlusion comes in pairs between consecutive bounces, so we consider vertices and segments separately. Second, it is possible that an incident ray reaches a microfacet while coming from lower hemisphere of the macro surface. Thus, we will need a fully-spherical formulation of the shadowing-masking functions, instead of the usual hemi-spherical formulation. Heitz et al.~\shortcite{heitz2016} mentioned the full sphere definition for VNDF, but did not point it out explicitly. We will elaborate it after the introduction of our path formulation.

\subsection{Position-free path integral}

We define the light transport at any shading point $\bs$, potentially undergoing multiple bounces, as a path integral for a given pair of query directions $\omega_i$ and $\omega_o$. 

The light might bounce several times before exiting the microsurface, and we define each bounce as a \textbf{vertex} $b_i$. We treat the position of the vertices as identical and focus on the two adjacent directions, rather than positions or depth. This makes our position-free path formulation completely independent of positions, even simpler than for layered materials which requires a depth to be recorded and is only position-free ``horizontally''.

\myfigure{path}{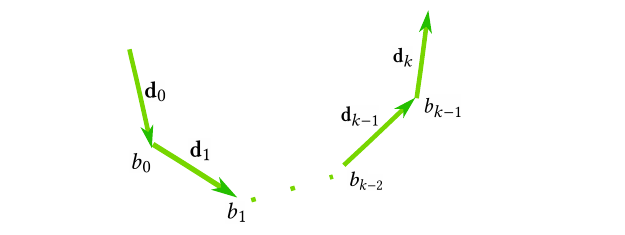}{ A light path is defined by $\bar x = ({\bd_0,b_0,\bd_1,b_1,\dots,b_{k-1},\bd_k})$.}

A \textbf{direction} is mostly the same as that in macro scale. We use $\bd_i$ as a unit vector on $\mathcal{S}^{2}$ to denote the light bouncing among the microfacets. The only difference is that we use the natural flow of light, i.e., assuming the incident pointing inwards instead of outwards any vertex $b_i$.

Now we define a \textbf{light path} $\bar x$ as a sequence of vertices and directions: $\bar x = ({\bd_0,b_0,\bd_1,b_1,\dots,b_{k-1},\bd_k})$, as shown in Figure~\ref{fig:path}. The first and last directions are aligned with the macro incident and outgoing directions of a BSDF query, i.e., $\bd_0=-\omega_i$ and $\bd_k=\omega_o$.

The \textbf{path contribution $f(\bar x)$} of a light path is the product of vertex terms $v_i$ (on each vertex) and segment terms $s_i$ (on each direction):
\begin{equation}
f(\bar x) = s_0 v_{0} s_1 v_{1} \cdots v_{k-1}s_{k}.
\label{eq:path}
\end{equation}

Based on our earlier analysis, we define the \textbf{vertex term} $v_i$ to represent local interactions between the light and the microfacets. It consists of everything except the shadowing-masking term, i.e., the normal distribution function $D$, the Fresnel term $F$ and the Jacobian term together:
\begin{equation}
v_i =  \frac{F(-\bd_i, \bomega_h^{i})\ D(\bomega_h)}{4 \vert \bomega_h^{i} \cdot (-\bd_i) \vert \vert \bomega_h^{i} \cdot \bd_{i+1} \vert},
\end{equation}
where $\bomega_h^{i}$ is the half vector between $-\bd_i$ and $\bd_{i+1}$.

The \textbf{segment term} $s_i$ describes the amount of energy leaving the previous vertex (if any) and arriving at the next (if any). We formulate it in detail in the next subsection. 

The path space $\Omega(\bomega_i,\bomega_o)$ is the set of all possible paths with their first directions equal to $-\bomega_i$ and the last directions equal to $\bomega_o$. Denoting the length of a path $\bar x$ as the number of directions in this path, a subspace $\Omega_{k}$ is then  the set of all possible paths with the same length $k$, and we immediately have $\Omega = \cup_{k\geq2}\Omega_{k}$.

The path space measure $\mu(\bar x)$ is a product of solid angle measures $\sigma$ at all vertices along the path toward their outgoing directions. That is, for a path with length $k$, 
\begin{equation}
\mu(\bar x) = \prod_{i=0}^{k-1}\sigma(d_i).
\end{equation}

Note that there is no measure for vertices, thanks to our position-free formulation.

Finally, we define \textbf{the multiple-bounce BSDF $\rho(\omega_i,\omega_o)$} as an integral over the set of paths $\Omega(\omega_i,\omega_o)$:
\begin{equation}
\rho(\omega_i,\omega_o) =  \int_{\Omega(\omega_i,\omega_o)}{f(\bar x) \dd\mu(\bar x)}.
\label{eq:pathint}
\end{equation}

\subsection{Full-spherical segment term and shadowing-masking term}

\myfigure{segmentterm}{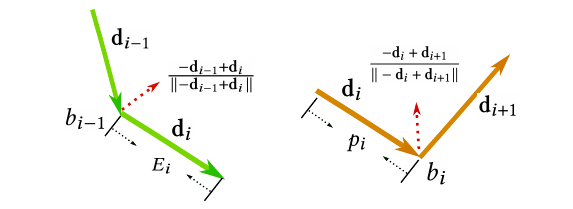}{ The segment term $s_i$ includes two different parts: $E_{i}$ and $p_{i}$, where $E_{i}$ represents the exit probability for $\bd_i$ at bounce $i-1$ and $p_{i}$ accounts for the effect of occluding microfacets preventing the direction $\bd{i}$ from hitting the next vertex $b_{i+1}$ at bounce $i$.}

The meaning of the segment term $s_i$ is intuitive. It tells the outgoing directional energy distribution at vertex $b_i$ along direction $\bd_{i+1}$. But before receiving energy at the next bounce, we need to ensure that the reflected / refracted light will participate in the next bounce. We formulate the segment term into two different parts (Fig.~\ref{fig:segmentterm}) as
\begin{equation}
s_i = e_{i} \cdot p_{i}.
\end{equation}

The first part $e_{i}$ is the exit probability. When the light bounces on vertex $b_{i-1}$, suppose there is no shadowing and masking, $100\%$ of the reflected / refracted energy towards $\bd_{i}$ (given by the vertex term $v_{i-1}$) will remain unoccluded as the light leaves the vertex $b_{i-1}$. But with potential shadowing and masking, part of the energy will be occluded as the light exits, inflicting the next bounce, while the other part of the energy will never touch the microstructure again, thus stopping further bounces of the light.

We mathematically define the exit probability as 
\begin{equation}
	e_i^{(0<i<k)} = \left\{
	\begin{array}{lr} 
	    1-G_1\left(\bd_{i},\frac{-\bd_{i-1} + \bd_{i}}{\Vert-\bd_{i-1} + \bd_{i}\Vert}\right), \mathrm{~~if~~} \bd_{i} \cdot \omega_g > 0 \\
	    1, \mathrm{~~if~~} \bd_{i} \cdot \omega_g \leq 0, 
    \end{array}  
	\right.
	\label{eq:E}
\end{equation}
where $G_1$ is the usual single-sided shadowing-masking term~\cite{smith:1967:smith}, the proportion of microfacets that are not occluded from a given direction. Also, when the light bounces downwards the macro surface, the next bounce is always guaranteed to happen, as the macro surface is watertight.

There are two special cases -- the first and the last segments. For the first segment, since it does not have to exit any previous vertex, its value is always $e_0 = 1$. The other special case on the last bounce is easily understood, as we would like to continue the next bounce for all other vertices except the last one, where we actually need the path to stop bouncing to keep its total length to $k$. Therefore, the exit probability becomes the ``inverse'' of others as
\begin{equation}
	e_i^{(i=k)} = \left\{
	\begin{array}{lr} 
	    G_1(\bd_{i},\frac{-\bd_{i-1} + \bd_{i}}{\Vert-\bd_{i-1} + \bd_{i}\Vert}), \mathrm{~~if~~} \bd_{i} \cdot \omega_g > 0 \\
	    0, \mathrm{~~if~~} \bd_{i} \cdot \omega_g \leq 0, 
    \end{array}  
	\right.
\end{equation}

\myfigure{sph_motivation}{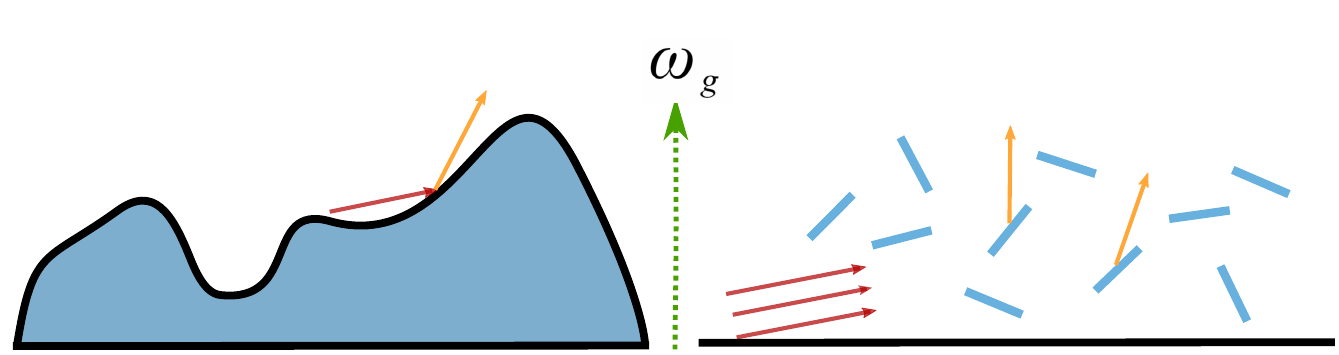}{Under the microfacet theory, it is possible that an incident ray coming from the lower hemisphere (the other side of the macro normal) be reflected, either by the actual microsurfaces (left), or by the microflakes in the Smith theory (right).}

Now we take a look at the second part $p_{i}$, which accounts for the effect of occluding microfac
ets preventing the direction $\bd_{i}$ from hitting the next vertex $b_{i+1}$ (if any). From the definition, we immediately know that this is the single-sided shadowing-masking term $G_1$. However, in this case, we must explicitly deal with the possible incident direction $\bd_{i}$ from below the macro surface to the vertex $b_{i+1}$, as seen in Fig.~\ref{fig:sph_motivation}.

The full-spherical shadowing-masking term $G_1$ was implicitly indicated by Heitz et al.~\shortcite{heitz2016} as the absolute value of its original hemispherical version. But to our knowledge, there is no explicit explanation or derivation available. Therefore, we provide a detailed derivation in the supplemental materials, and provide our conclusion: 
\begin{equation}
	G_1(\omega,\omega_m) = G_1^\mathrm{local}(\omega,\omega_m) G_1^\mathrm{dist}(\omega), 
\end{equation}
where $G_1^\mathrm{local}(\omega,\omega_m)=\chi^{+}(\omega\cdot{\omega_m})$, which is $1$ when $\omega$ and $\omega_m$ are facing in the same direction, i.e., $\omega \cdot \omega_m>0$, and is $0$ otherwise. $G_1^\mathrm{dist}$ is the distant shadowing / masking term, in our full-spherical case:
\begin{equation}
G_{1}^\mathrm{dist}(\omega) =  \left| \frac{1}{1 + \Lambda(\omega)} \right| = \left\{
\begin{array}{lr} 
	1 / (1 + \Lambda(\omega)), \mathrm{~~if~~} \omega \cdot \omega_g > 0, \\
	-1 / (1 + \Lambda(\omega)), \mathrm{~~if~~} \omega \cdot \omega_g \leq 0, 
\end{array}
\right.
\label{eqn:g1distABS:2}
\end{equation}

$\Lambda(\omega)$ is the Smith Lambda function, which is analytical for both Beckmann and GGX models. More details could be found in Heitz et al.~\shortcite{heitz2016}.

With the explicit full-spherical shadowing-masking function, we are able to define $p_i$ as  
\begin{equation}
	p_i^{(i<k)} = G_1\left(-\bd_{i},\frac{-\bd_{i} + \bd_{i+1}}{\Vert-\bd_{i} + \bd_{i+1}\Vert}\right), 
\end{equation}
and	$p_k=1$ since the last segment exiting the surface will not hit any more vertices.

So far, we have the complete segment term $s_i$ derived. Note that there is no double counting of occlusion along the same direction $\bd_{i+1}$, as $e_i$ is for exiting vertex $b_i$ and $p_{i+1}$ is for entering vertex $b_{i+1}$. They are different vertices, even with our position-free formulation.

\subsection{Properties and analysis}
\label{sec:analysis}
Now that we have a complete multiple-bounce BSDF formulation, we briefly analyze it to check that it has the right properties.

\paragraph{Position-free.} Our BSDF formulation is completely independent of the positions of individual vertices. This immediately demonstrates that both in the macro scale and in the micro scale, our method is position-free. Therefore, there is also no need to keep track of the height of a vertex as in~\cite{heitz2016}.

\paragraph{Height correlation.} Since we treat the vertex terms and segment terms separately, our method is essentially implying a height-uncorr- elated shadowing-masking model. This enables a clean and explicit formulation for multiple bounce computation, and does not prevent our model from passing the white furnace test both in theory and in practice (Fig.~\ref{fig:glassRoughness}). Therefore, while incorporating height correlation in our vertex and segment terms may result in further enhancements, we leave this improvement for future work.

\paragraph{Generality.} One can easily verify the generality of our path formulation in Eqn.~\ref{eq:pathint}. It reduces to classic single-bounce BSDFs, if we limit the length of paths to $2$. Therefore, our formulation is a general definition of BSDFs.

\paragraph{Reciprocity.} From the structure of the path integral, we can see that its overall reciprocity lies in the individual vertex terms and segment terms. The reciprocity of the vertex terms is trivial to verify, since they are essentially traditional microfacet BSDFs without the shadowing-masking terms. In the supplemental materials, we prove that the segment terms are also reciprocal, and provide an experiment to demonstrate this. Therefore, our entire BSDF formulation is reciprocal. 

\paragraph{Normal mapping support.} In Schussler et al.~\shortcite{Schussler:2017:normal}, it is pointed out that regular normal mapping on hemispherical BSDFs will inevitably confuse the sides of incident and outgoing directions, leading to black regions when the specified normals deviate much from the original. However, since our BSDF formulation is fully spherical, directly applying normal mapping will never cause similar issues. Therefore, no additional effort needs to be done to support correct normal mapping. We demonstrate this in Fig.~\ref{fig:normalmap}.

\paragraph{Variance reduction.} As mentioned before, our explicit path integration enables any Monte Carlo solutions to it. This property allows us to introduce much more efficient estimators than previous random walk methods, which reduces the variance significantly, as we show next in Sec.~\ref{sec:pathint}. Note that Heitz et al.~\shortcite{heitz2016} use multiple importance sampling (MIS) to combine the contribution from two random walk paths, one pure forward and the other backward. Our explicit formulation is different from theirs and allows full bidirectional approaches, enabling connection of half paths from both directions, reusing much more samples and resulting in less variance.

%-----------------------------------%
\section{Monte Carlo Path Integral Estimators}
\label{sec:pathint}
%-----------------------------------%

With our explicit and position-free path formulation, any Monte Carlo method can be used to compute the integral. In this section, we propose two estimators for BSDF evaluation: unidirectional path tracing (PT) and bidirectional path tracing (BDPT) to evaluate the multiple scattering path integral, inspired by the position-free integral that solves the BSDFs of layered materials~\cite{Guo:2018:Layered}. We show how to efficiently sample our multiple-bounce BSDFs, and the computation of the corresponding probability density functions. Throughout this section, we use conductors as examples, without loss of generality.

\subsection{Path Tracing} 
\label{sec:pathint:pt}

We first propose a unidirectional estimator using path tracing for BSDF evaluation: 
\begin{equation}
\rho(\omega_i,\omega_o) \approx  \frac{1}{N}\sum_{j=0}^{j=N} \frac{f(\bar x)}{\mathrm{pdf}(\bar x)},
\end{equation}
where $N$ is the sample count, $\bar x$ is a sampled path starting from $\bd_0$ and $\mathrm{pdf}(\bar x)$ is the probability density function (pdf) of the sample path. We set N as 1 for each BSDF evaluation. Since the path has to be ended with $\omega_o$, it's impossible to reach such a direction with directional sampling only, thus we perform the next event estimation (NEE) from the final outgoing direction $\omega_o$ for each bounce, resulting in:
\begin{equation}
\rho(\omega_i,\omega_o) \approx  \frac{1}{N}\sum_{j=0}^{j=N} \sum_{i=2}^{i=k+1} \frac{f(\bar x_{i})}{\mathrm{pdf}(\bar x_{i})},
\end{equation}
where $\bar x_{i} = (\bd_0,b_0,\bd_1,b_1,\dots,b_{i-2},\bd_{i-1})$ represents a path with length $i$ from path space $\Omega_{i}$ and $f(\bar x_{i})$ is computed with Equation~\ref{eq:path}. The pdf of a path $\mathrm{pdf}(\bar x_{i})$ is computed as the product of all the pdf to sample the internal directions, from $\bd_1$ to $\bd_{i-2}$:
\begin{equation}
\mathrm{pdf}(\bar x_i) = \prod_{j=0}^{j=i-3}\mathrm{pdf}(-\bd_j,\bd_{j+1}), 
\end{equation}
where $\mathrm{pdf}(-\bd_j,\bd_{j+1})$ represents the pdf of sampling $\bd_{j+1}$ from $-\bd_{j}$. Note that both $f(\bar x_{i})$ and $\mathrm{pdf}(\bar x_{i})$ are evaluated recursively. 

For the specific sampling method and corresponding pdf values, we simply refer to the distribution of visible normals (VNDF) sampling technique~\cite{Heitz2014imporSAM}.

Finally, note that our path tracing method provides an unbiased estimation of the multiple-bounce BSDF with \emph{unlimited number of bounces}. This is in essence different from the multiple-bounce BSDFs under the V-groove assumption~\cite{lEE2018PRATMUL}, in which a maximum bounce must be specified, balancing between potential energy cutoff and computational overhead. In theory, we do not have to set the maximum-bounce. However, in practice, we set it to $10$ for simpler implementation and better performance. We do not currently use Russian roulette, but it would be easy to enable it.

\subsection{Bidirectional path tracing}
\label{sec:path:bdpt}

We now present an even more efficient bidirectional estimator, following the classical bi-directional approach in light transport. We first trace rays from both $\bd_0$ and $-\bd_k$ with maximum length $k + 1$ and generate a camera path and a light path. Then we combine them, choosing $s$ directions from camera path, and $t$ directions from the light path, where $ 0 < s < k + 1$ and $ 0 < t < k + 2 - s $. If $s=1$, then only $\bd_0$ is chosen from the camera path. Similarly, if $t=1$, then only $\bd_k$ is taken from the light path. For each generated path $\bar x_{i}$, we compute its contribution $f(\bar x_{i})$, $\mathrm{pdf}(\bar x_{i})$, and the MIS weight $w(\bar x_{i})$.

The path contribution $f(\bar x_{i})$ of path $\bar x_{i}$ with length $i$ is computed with Equation~\ref{eq:path}, by accumulating all the vertex terms and the segment terms along the path.

The pdf is computed by accumulating the $s$ pdfs from the camera path and $t$ pdfs from the light path:
\begin{equation}
	\mathrm{pdf}(\bar x_{i}) = \prod_{j=1}^{j=s} \mathrm{pdf}(-\bd_{j-1},\bd_j) \prod_{j=k - s}^{j = k } \mathrm{pdf}(\bd_j,-\bd_{j-1}),
	\label{equ:pdf}
\end{equation}
Note that, we start from 1 rather than 0, as the both $\bd_0$ and $\bd_k$ are not sampled.

Regarding the MIS weight, for a given path $\bar x_{i}$ with length $i$, there are $i-1$ possible ways to generate this length, by taking different number of directions from the camera path and the light path. We sum up all the pdf for each possible way as $\sum{\bar x_{i}^j}$, and then compute the MIS weight with the balance heuristic, as:
\begin{equation}
	w(\bar x_{i}) = \frac{\mathrm{pdf}(\bar x_{i})}{\sum{\mathrm{pdf}(\bar x_{i}^j)}}.
	\label{equ:misweights}
\end{equation}

Finally, we get the bidirectional estimator for multiple-bounce BSDF as:
\begin{equation}
	\rho(\omega_i,\omega_o) = \sum_{i=2}^{i=k+1} \frac{f(\bar x_{i}) w(\bar x_{i})}{\mathrm{pdf}(\bar x_{i})}.
\end{equation}

The bidirectional estimator produces results with less variance, since there are implicitly more paths used for estimation. The paths are also weighted in a proper way, which allows to further reduce variance.

Also, from the definition of paths in our formulation, we can see that they are completely consisted with directions, thus is completely position-free. This is different from Guo et al.~\shortcite{Guo:2018:Layered}, where the position-independence is only in the horizontal direction, while they keep trace of the depths into the different layers for their path formulations.

\subsection{Importance sampling and PDF of multiple bounces}
\label{sec:path:sampling}

Importance sampling is required to fit our BSDF in a path tracing framework. It's straight forward to do sampling in our BSDF. For a given incident direction $\omega_i$, the sample function should answer the final $\omega_o$ and compute the sampling weights.

Let's consider bounce $i$. Starting from $\bd_i$, we use the VNDF importance sampling to generate an outgoing direction $\bd_{i+1}$. With this sampled direction, we compute the Fresnel factor and use it as a probability to decide reflecting or refracting the ray. Then we check whether $\bd_{i+1}$ points towards the macro surface. If so, we continue sampling. Otherwise, we compute the masking function $G_1(\bd_{k+1})$ and use it as a probability to choose leaving the surface or performing more bounces. If choosing to leave the surface, $\omega_o$ is obtained. If the maximum bounces are reached, but the ray has not left the surface, the sampling fails, which is usual in the light transport.

For conductor material, the sampling weights include all the Fresnel factor along the path, since all the other terms are canceled out by the VNDF sampling and the exit surface sampling. For a rough dielectric BSDF, the weight is 1, since all the terms are canceled out. locate, we 

The pdf function of a BSDF is used in MIS. Given a $\omega_i$ and $\omega_o$, it should answer the pdf for this setting. Since we are using a random walk in our BSDF, it's impossible to get the exact pdf for a $\omega_i$ and $\omega_o$ pair. Hence, we use the same method as Heitz et al.~\shortcite{heitz2016}, combining the pdf of single-scattering and a diffuse term pdf.

\myfigure{normalmap}{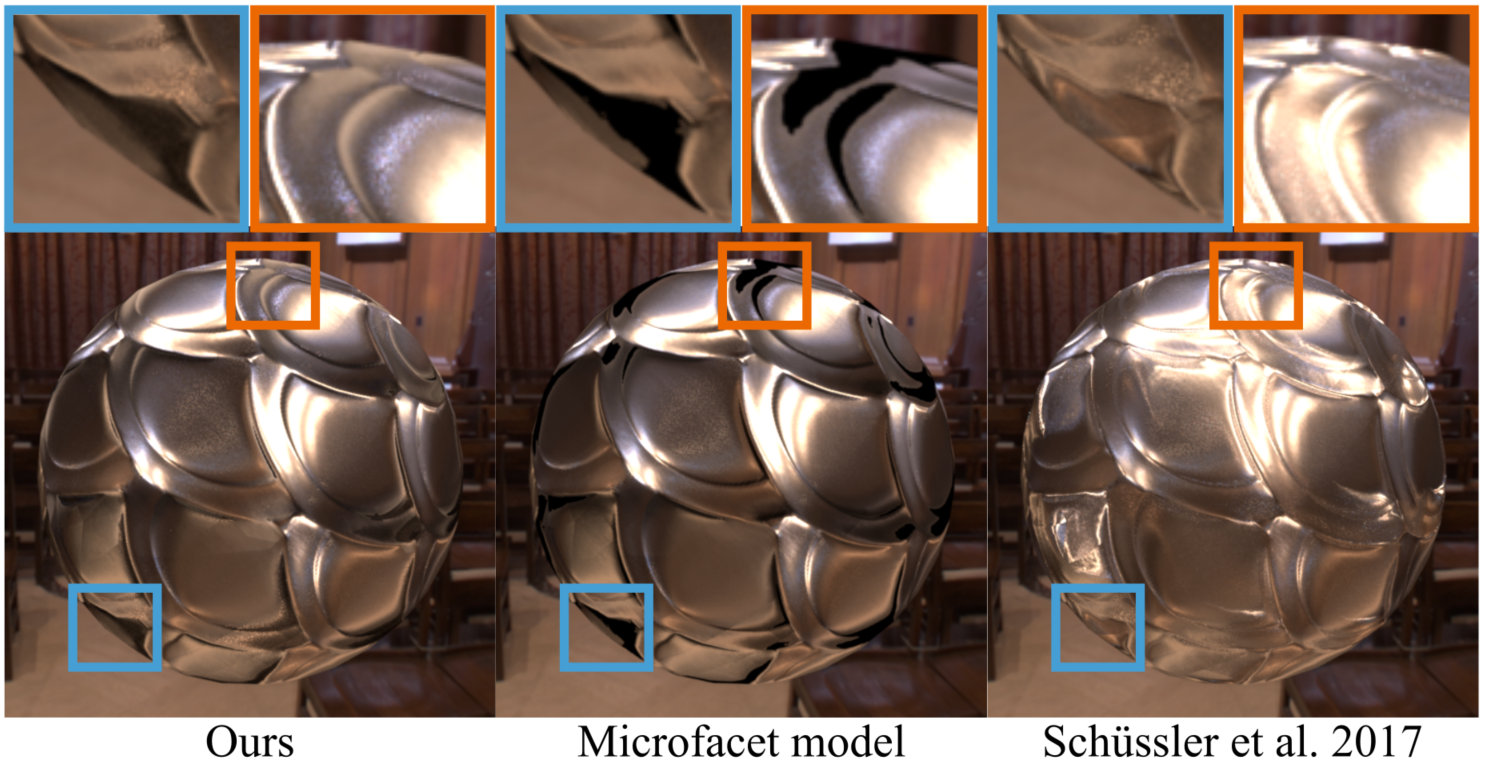}{Comparison between our method, Sch\"{u}ssler et al.~\shortcite{Schussler:2017:normal} and the microfacet model on a metal sphere with a normal map. Our method does not have the black artifacts in microfacet models. Since Sch\"{u}ssler et al.~\shortcite{Schussler:2017:normal} use a different model, it produces a different result from ours, as expected.}

\myfigure{conductorBDPT}{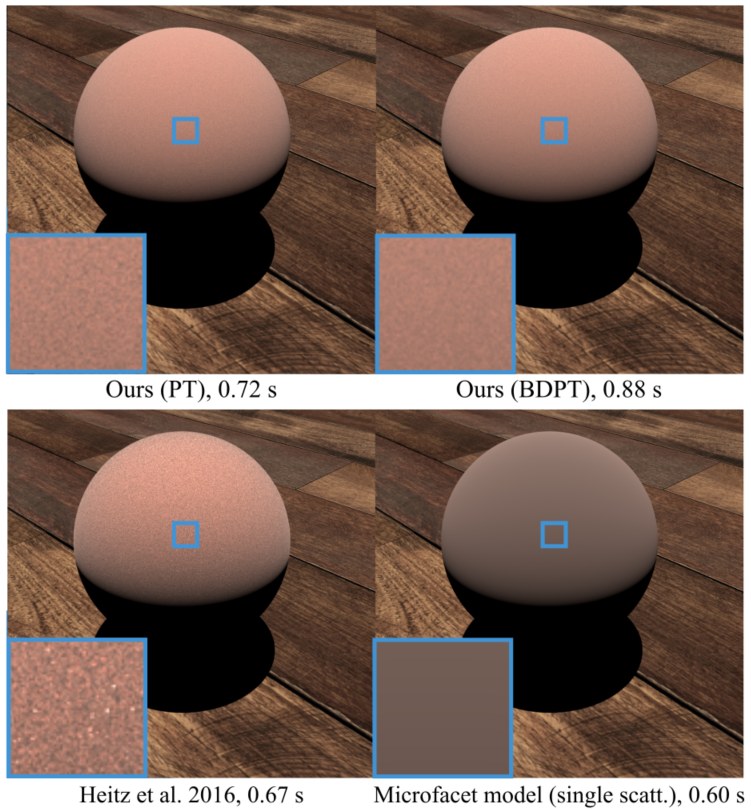}{Comparison between our method (PT), our method (BDPT), Heitz et al.~\shortcite{heitz2016} and the classical microfacet model on a rough conductor BSDF (GGX model, $\alpha$ = 1.0) with 4 spp. Both of our unidirectional and bidirectional methods produce less noise than Heitz et al.~\shortcite{heitz2016}, while our bidirectional approach produces higher quality result than our path tracing approach, with some extra time cost.}

\myfigure{convergence}{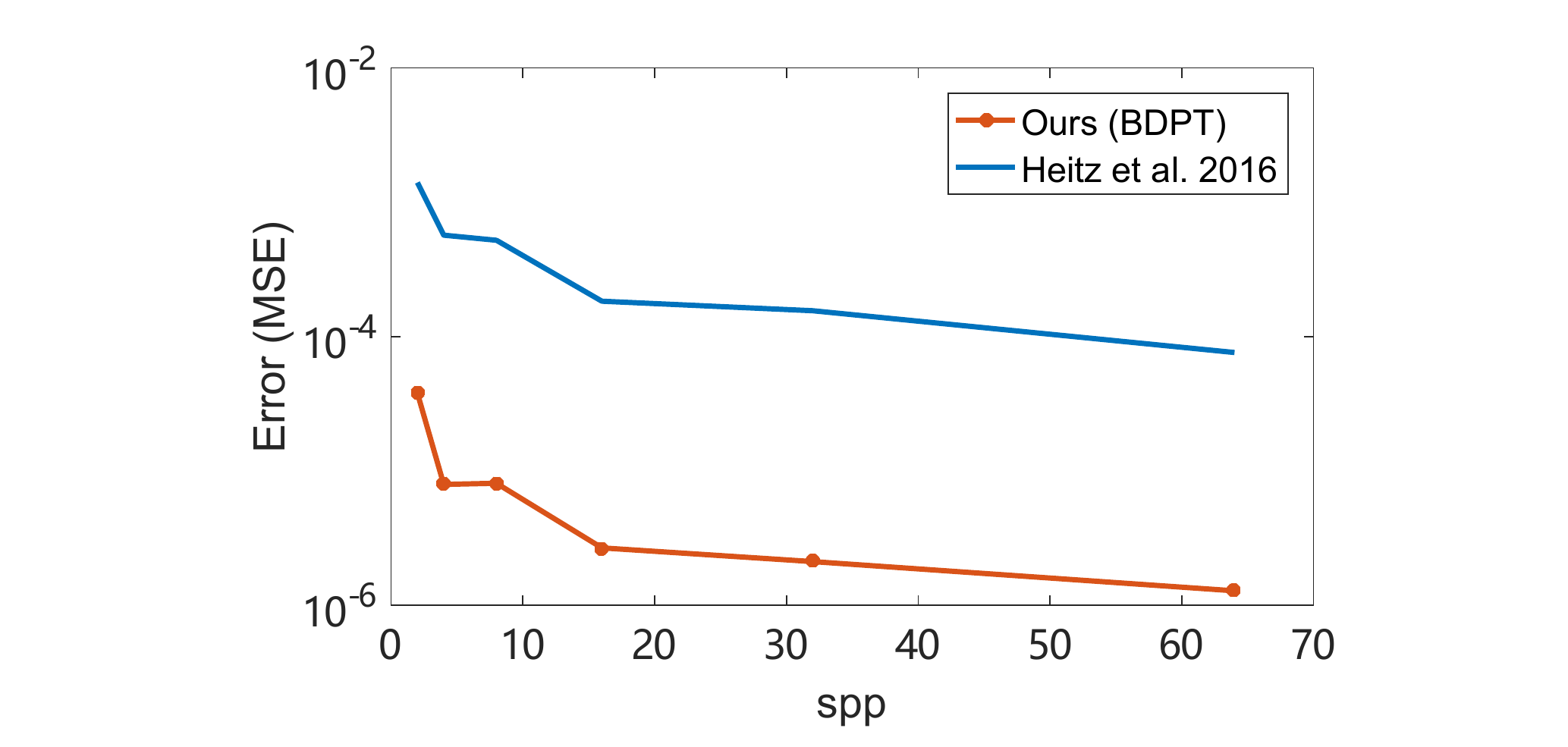}{The error (MSE) with logarithm scale of our method and Heitz et al.~\shortcite{heitz2016} over varying spp on the Single Slab scene (Figure~\ref{teaser}(c) and (d)).}

\mycfigure{kitchenVase}{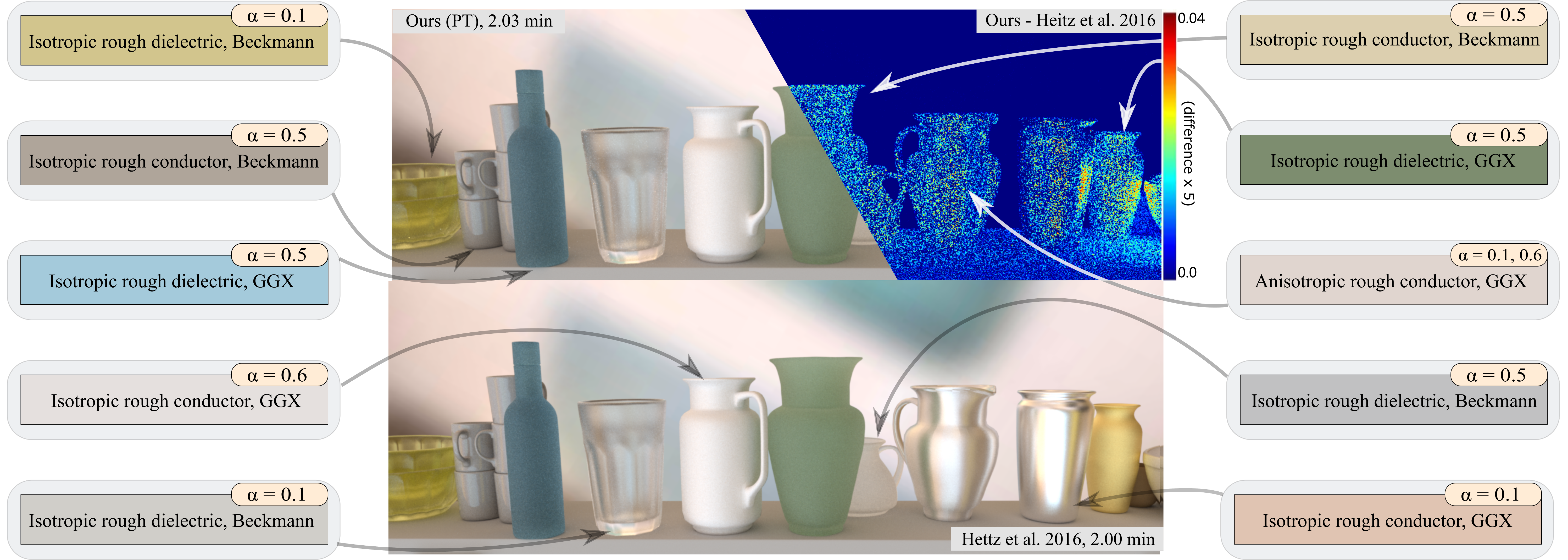}{Our result (rendered with PT) and Heitz et al.~\shortcite{heitz2016} on different materials at roughly equal time. The material settings are shown in the image. We visualize the difference (5$\times$ scaled) between ours and Heitz et al.~\shortcite{heitz2016}.}

\mycfigure{glassRoughness}{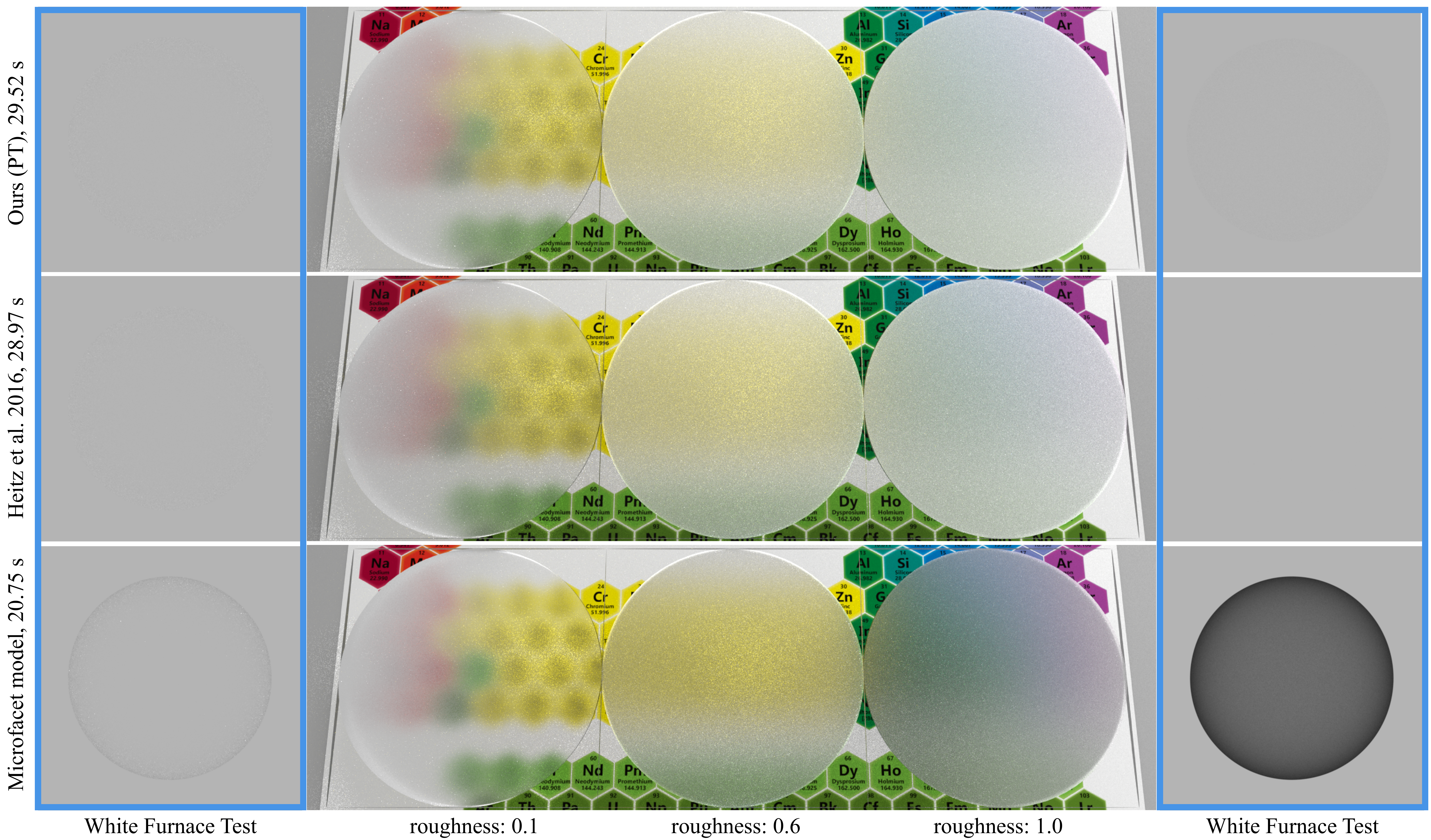}{Comparison between our method (PT), Heitz et al.~\shortcite{heitz2016} and microfacet model on rough dielectric BSDFs (GGX model) with varying roughness (0,1, 0.6 and 1.0), rendered with 64 spp. Our unidirectional estimator produces results very similar to Heitz et al.~\shortcite{heitz2016} and has almost identical computation time. The left and right most images show the white furnace test results of BSDFs with roughness 0.1 and 1.0 respectively. }

\mycfigure{svRoughness}{svRoughness.pdf}{Comparison between our method (PT) and microfacet model on rough dielectric BSDFs (GGX model) with spatial-varying roughness.}

\myfigure{vGroove}{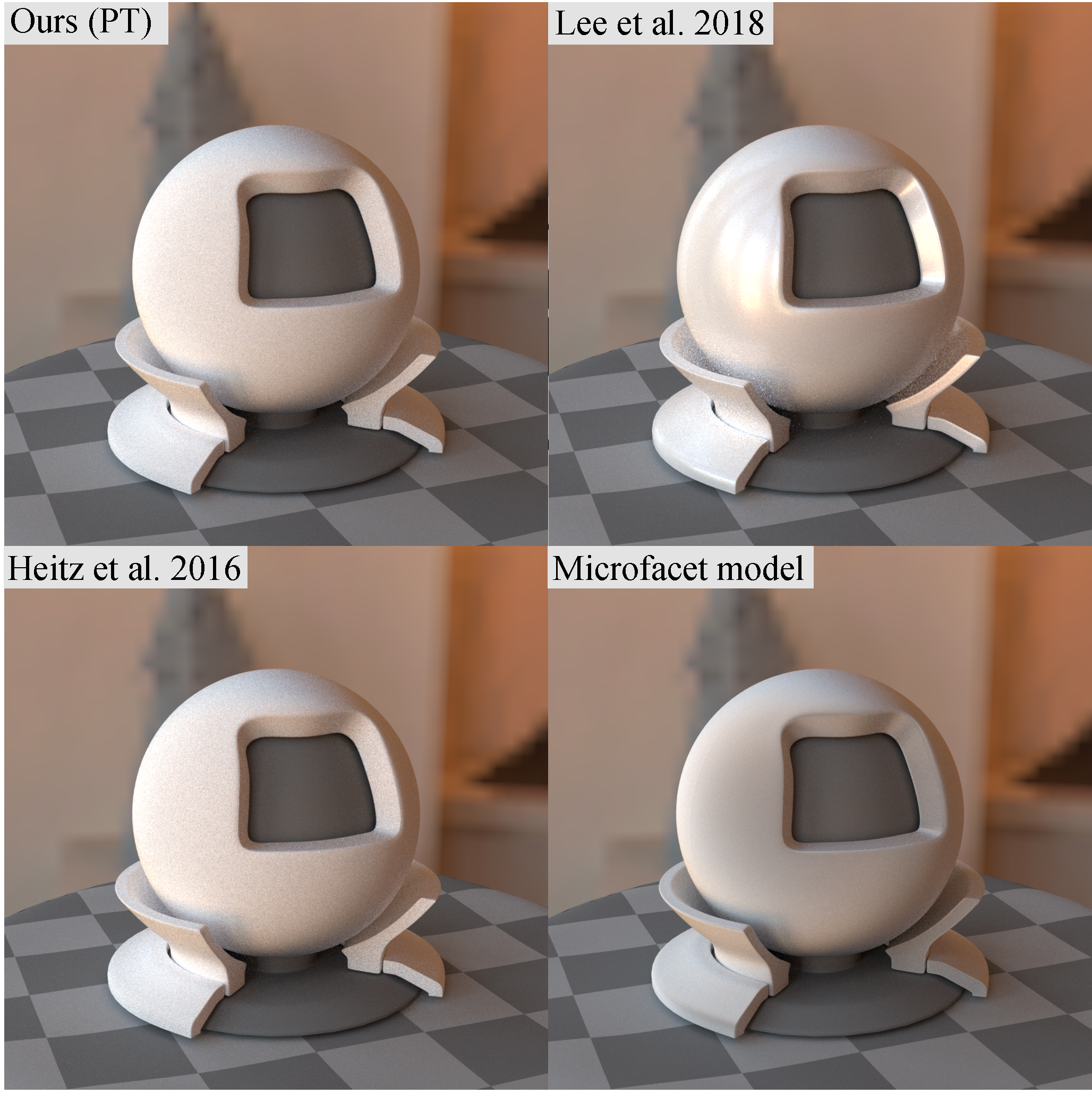}{Comparison between our method (PT), Heitz et al.~\shortcite{heitz2016} and Lee et al.~\shortcite{lEE2018PRATMUL} for a rough conductor BSDF (Beckmann model, $\alpha$ = 1.0). The result for Lee et al.~\shortcite{lEE2018PRATMUL} is too glossy; this issue is a drawback of the V-groove shadowing model.}

\myfigure{kulla}{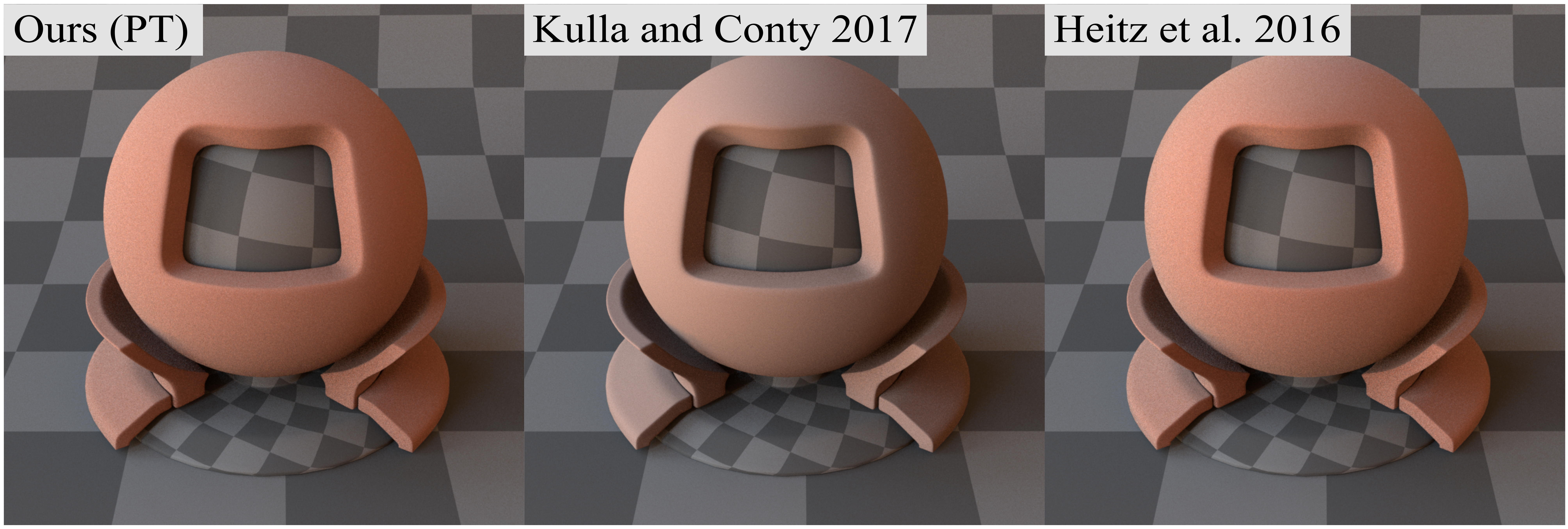}{Comparison between our method (PT), Kulla and Conty~\shortcite{KullaConty:2017:revisiting} and Heitz et al.~\shortcite{heitz2016} for a rough conductor BSDF (GGX model, $\alpha$ = 1.0). Kulla and Conty~\shortcite{KullaConty:2017:revisiting} average the Fresnel term over all directions, resulting in a significantly different color. Both our method and Heitz et al.~\shortcite{heitz2016} compute the Fresnel term at each bounce.}

\mycfigure{UEResult}{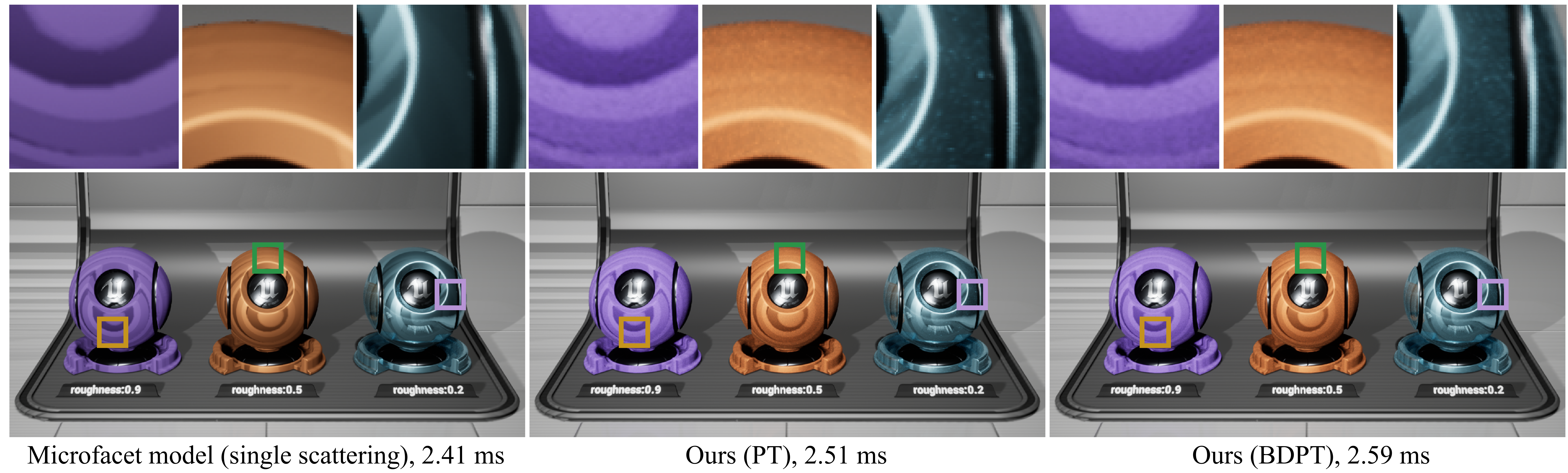}{Comparison between the microfacet model (single scattering only) and our methods (PT and BDPT, both 1 spp) implemented in shaders in UE4 on three material balls with different roughness ($0.9$, $0.5$ and $0.2$). Our methods (PT and BDPT) bring obvious color changes, especially when roughness is high. With DLSS that is readily available in UE4, even using only 1 spp, both our PT and BDPT methods achieve low noise level, and our BDPT is almost noise-free for all three materials.}

\section{Results and Comparison}
\label{sec:results}

We have implemented our algorithm inside the Mitsuba renderer \shortcite{Mitsuba} for both rough conductor and rough dielectric BSDFs. The implementation of Heitz et al.~\shortcite{heitz2016} is from the author's website, with bidirectional random walk. All timings in this section are measured on a 2.20GHz Intel i7 (40 cores) with 32 GB of main memory. For the reference images, we simply refer to the converged result using the same method. This is because in theory there is no ground truth, and Heitz et al.~\shortcite{heitz2016} and our method converge to different results. Therefore, to have a fair estimation of noise level, we compare different methods with their own converged results as references.

\paragraph{Comparison of lobes for individual bounces} 
In the supplemental materials, we compare the visualized lobes for individual bounce between our method and Heitz et al.~\shortcite{heitz2016}. We perform the comparison on rough conductor (Fresnel set as 1) and rough dielectric BSDFs, considering both isotropic ($\alpha = 1$) and anisotropic ($\alpha = (0.1, 1.0)$) cases. We visualize the lobes with $\omega_i$ elevation angles of 0.0 and 1.5 radians. $E_r$ and $E_t$ denote the total amount of reflected and transmitted energies, respectively. The differences between our method and Heitz et al.~\shortcite{heitz2016}'s appear mostly at grazing angles, as we use the height-uncorrelated shadowing-masking function, while they used the height-correlated one. Also, since both methods pass the white furnace test (thus are both correct and energy conserving), it is possible that our method is brighter than Heitz et al.~\shortcite{heitz2016} at some particular bounces, but it cannot be consistently brighter for all bounces.

\paragraph{Evaluation-only comparison.} 
Our method is especially suitable to render under sharp lighting and in evaluation-heavy situations, thus we first show some results with direct lighting only. In Fig.~\ref{fig:conductorBDPT}, we show a copper sphere (GGX model, $\alpha$ = 1.0) lit by a directional light. Both of our PT and BDPT methods produce much less noise than Heitz et al.~\shortcite{heitz2016}, with a slight performance overhead, and our BDPT approach produces the best result. Fig.~\ref{teaser} (c) and (d) show a single slab with a rough dielectric BSDF (GGX model, $\alpha$ = 1.0). Both our path tracing and our BDPT produce better results than Heitz et al.~\shortcite{heitz2016}, while our path tracing method is faster than Heitz et al.~\shortcite{heitz2016} and our BDPT reduces the noise significantly with acceptable extra time cost. With only two samples per pixel, our method (BDPT) produces results close to noise-free. In Figure~\ref{fig:convergence}, we show the Mean Square Error (MSE) as a function of varying spp for our method (BDPT) and Heitz et al.~\shortcite{heitz2016} in the Single Slab scene, considering directional lighting only. With only two samples per pixel, our method is able to produce very close result to the ground truth, while Heitz et al.~\shortcite{heitz2016} produces result with a lot of noise. Increasing the number of samples improves the quality for both methods, but our method remains consistently better. In the supplemental materials, we show more convergence comparisons between our method and Heitz et al.~\shortcite{heitz2016} over varying roughness. For all these configurations, our method shows better quality.

\paragraph{Equal-time comparison.} 
In Fig.~\ref{teaser}(a) and (b), we show three deer statues (copper (GGX model, $\alpha$ = 0.1), aluminum (GGX model, $\alpha$ = 0.6) and gold (GGX model, $\alpha$ =0.5)) on an aluminum floor (GGX model, $\alpha$ = 0.1), lit by an environment map and a point light. To better show the effect of the BSDF evaluation, we use 64 spp for the environment map lighting. 
To achieve equal time, we use 8 spp for our method and 14 spp for Heitz et al.~\shortcite{heitz2016} for the point light source. Our results have much less noise than Heitz et al.~\shortcite{heitz2016}'s result. We also report the MSE of the entire image, which confirms the high quality of our results. 

\paragraph{Complex lighting comparison.} 
In Fig.\ref{fig:kitchenVase}, we compare against Heitz et al.~\shortcite{heitz2016} on more complex lighting, considering both indirect illumination and environment lighting. Our results are almost identical to Heitz et al.~\shortcite{heitz2016}. In this scene, our method does not significantly reduce the noise. There are two reasons: first, in this scene, light transport is complex and responsible for most of the noise; second, our method for sampling BSDFs produces similar noise as Heitz et al.~\shortcite{heitz2016}. Hence, our method is especially suitable to render under sharp lighting and in evaluation-heavy situations.

In Fig.~\ref{fig:glassRoughness}, we show the results of our method (PT), Heitz et al.~\shortcite{heitz2016} and single-bounce microfacet model for rough dielectric BSDFs with varying roughness. Our unidirectional estimator produces results very similar to Heitz et al.~\shortcite{heitz2016} and has almost identical costs. We also show the white furnace test results of the three methods, by rendering these materials lit by a constant white environment map. Both of our method and Heitz et al.~\shortcite{heitz2016} pass the white furnace test. In Fig.~\ref{fig:svRoughness}, we show the result of our method and microfacet model with spatial-varying roughness. As expected, microfacet model misses energy on materials with large roughness.

\paragraph{Comparison with other methods.} 
In Fig.~\ref{fig:vGroove}, we compare our method with Lee et al.~\shortcite{lEE2018PRATMUL} (nonsymmetric). The result of Lee et al.~\shortcite{lEE2018PRATMUL} inherits the drawback of V-groove approaches, and the results look too glossy. Both of our method and Heitz et al.~\shortcite{heitz2016} are based on the Smith shadowing method, and produce similar results. In Fig.~\ref{fig:kulla}, we compare our method with Kulla and Conty~\shortcite{KullaConty:2017:revisiting}. Their method is based on an average Fresnel term, instead of accumulating the Fresnel term contributions during multiple scattering. This results in a significant difference in color. Both our method and Heitz et al.~\shortcite{heitz2016} use the Fresnel term at each bounce, resulting in the correct color.

\paragraph{Real-time rendering }.
We implemented our method using shaders inside the Unreal Engine 4 (UE4), to show that it can be used for real-time rendering. We implemented both PT and BDPT versions, using a fixed sampling rate of 1 spp. We limit the number of bounces to two, as it already covers most of the energy. All timings in this section are measured on a NVIDIA RTX 2080 Ti graphics card. We exploit the modern rasterization pipeline, readily available in UE4, to generate high quality image sequences with the help of Deep Learning Super Sampling (DLSS).

Fig.~\ref{fig:UEResult} and the companion video show that both our PT and BDPT methods correctly produce the appearance from multiple-bounce BSDFs, and that our BDPT method is almost noise-free. The computation cost for our methods is only about 0.10 ms (for PT) and 0.19 ms (for BDPT), compared to single scattering only. Our method has potential applications in real-time rendering. We also provide a ShaderToy implementation of our method (showing both PT and BDPT) in the supplemental materials. We could not implement  Heitz et al.~\shortcite{heitz2016} in Unreal Engine 4 (UE4), due to its complexity. According to Figure~\ref{fig:convergence} and Figure 5 in the supplemental materials, even after 5 ms, the results from Heitz et al.~\shortcite{heitz2016} are still much noisier than our BDPT method.

\section{Discussion and limitations}

\paragraph{Shadowing-Masking function.}
The main reason for the differences between Heitz et al.~\shortcite{heitz2016} method and ours is the choice of the shadowing-masking function. We use the height uncorrelated shadowing-masking function, while Heitz et al.~\shortcite{heitz2016} use the height correlated version. As mentioned in Heitz~\shortcite{HEITZ2014UNDER}, the height correlated shadowing-masking function is also an approximation to the reality, since ``it overestimates shadowing when the directions are close''. Our position-free formulation (especially the definition of segment terms) naturally leads to the use of the height uncorrelated shadowing-masking functions, which introduces differences with Heitz et al.~\shortcite{heitz2016}, but results in less variance in our method by avoiding the need to trace the height and allowing bidirectional light transport.

\myfigure{limitation}{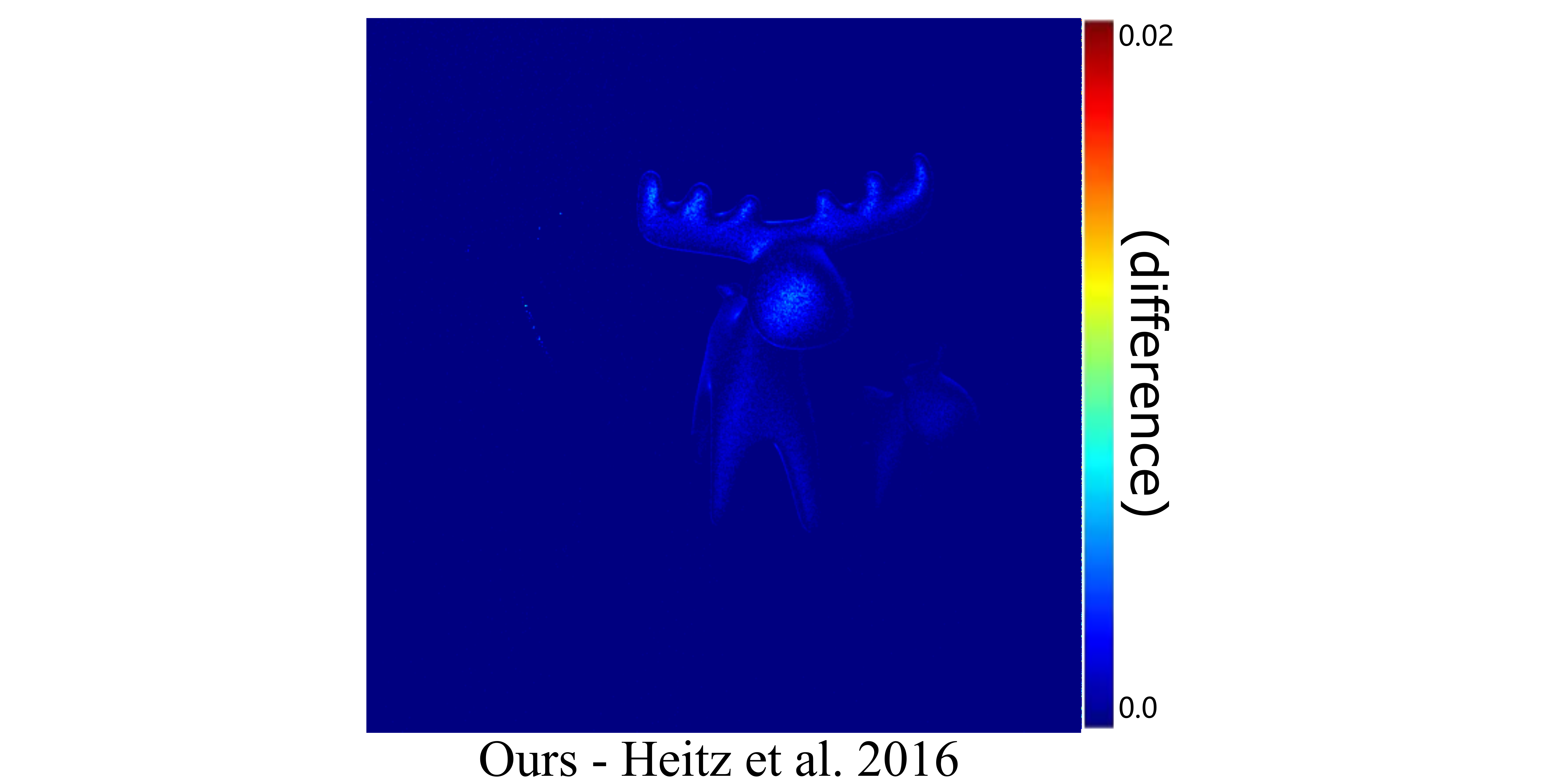}{Absolute differences between renderings using our method and Heitz et al.~\shortcite{heitz2016} (see Fig.~\ref{teaser} for the images). These differences are due to our use of a height-uncorrelated shadowing-masking term. Note again that both models can pass the white furnace test, only in different ways.}

Fig.~\ref{fig:limitation} shows the difference introduced by the use of height-correlated or height-uncorrelated shadowing-masking functions. In the supplemental, we also visualize the values of the shadowing-masking functions and BRDFs in the incident plane for both our method and Heitz et al.~\shortcite{heitz2016}.

\paragraph{Height distribution function.} Our method is independent from the choice of a height distribution function, since it is cancelled out during the computations on the $\Lambda$ function. In this respect, it is identical to Heitz et al.~\shortcite{heitz2016}. We haven't found existing previous work that studied the impact of different height distributions on multiple scattering; this would be an interesting venue for future work.

\paragraph{Limitations.} We have identified three main limitations for our method:
\begin{itemize}
  \item For performance reasons, our model currently focuses on NDFs with an analytical $\Lambda$ functions, such as Beckmann and GGX.
  \item Although it has a low variance, our method still does not have a closed-form solution. Finding an analytical solution would be a strong extension to this work.
  \item We use an height uncorrelated shadowing-masking function. Using a height-direction correlated shadowing masking function would improve accuracy, as mentioned in Heitz~\shortcite{HEITZ2014UNDER} , but it would require rewriting the path formulation.
\end{itemize}

\section{Conclusion and Future Work}
\label{sec:conclusion}
We have proposed a new formulation for multiple-bounce microfacet BSDFs under the Smith assumption. We start by deriving an explicit mathematical definition of the path space that describes single and multiple bounces in a uniform way and study the behavior of light on the different vertices and segments in path space. Then we pose the evaluation of our multiple-bounce BSDF as a position-free path integral and solve it with both path tracing and bidirectional path tracing. Our method produces much less noise than prior work, and is almost noise-free with very low sampling rate ($2-4$ spp) for BSDF evaluation tasks, thus is especially suitable to render under sharp lighting and in evaluation-heavy situations (scenes with small lights).

In the future, aside from eliminating the limitations of our method, it would be interesting to explore the possibility of applying our formulation on detailed surfaces with actual NDFs rather than statistical NDFs. Also, with our explicit formulation, coming up with a better sampling approach than the visible NDF sampling may also be useful to further reduce the variance. It is also worth looking into the possibilities to combine our method with precomputation-based methods, using pre-generated tables or textures to further speed up our run-time performance.

% \begin{acks}
% \end{acks}

%\newpage 
\bibliographystyle{ACM-Reference-Format}
\bibliography{paper}


\begin{thebibliography}{31}

%%% the Web address.

\ifx \showCODEN    \undefined \def \showCODEN     #1{\unskip}     \fi
\ifx \showDOI      \undefined \def \showDOI       #1{#1}\fi
\ifx \showISBNx    \undefined \def \showISBNx     #1{\unskip}     \fi
\ifx \showISBNxiii \undefined \def \showISBNxiii  #1{\unskip}     \fi
\ifx \showISSN     \undefined \def \showISSN      #1{\unskip}     \fi
\ifx \showLCCN     \undefined \def \showLCCN      #1{\unskip}     \fi
\ifx \shownote     \undefined \def \shownote      #1{#1}          \fi
\ifx \showarticletitle \undefined \def \showarticletitle #1{#1}   \fi
\ifx \showURL      \undefined \def \showURL       {\relax}        \fi
% invisible to TeX
\providecommand\bibfield[2]{#2}
\providecommand\bibinfo[2]{#2}
\providecommand\natexlab[1]{#1}
\providecommand\showeprint[2][]{arXiv:#2}

\bibitem[\protect\citeauthoryear{Ashikhmin and Premo\v{z}e}{Ashikhmin and
  Premo\v{z}e}{2007}]%
        {Ashikhmin:2007}
\bibfield{author}{\bibinfo{person}{Michael Ashikhmin} {and}
  \bibinfo{person}{Simon Premo\v{z}e}.} \bibinfo{year}{2007}\natexlab{}.
\newblock \bibinfo{title}{Distribution-Based BRDFs}.
\newblock
  \bibinfo{howpublished}{\url{https://citeseerx.ist.psu.edu/viewdoc/download?doi=10.1.1.214.8243&rep=rep1&type=pdf}}.
\newblock


\bibitem[\protect\citeauthoryear{Bagher, Soler, and Holzschuch}{Bagher
  et~al\mbox{.}}{2012}]%
        {bagher:2012}
\bibfield{author}{\bibinfo{person}{Mahdi~M. Bagher}, \bibinfo{person}{Cyril
  Soler}, {and} \bibinfo{person}{Nicolas Holzschuch}.}
  \bibinfo{year}{2012}\natexlab{}.
\newblock \showarticletitle{{Accurate fitting of measured reflectances using a
  Shifted Gamma micro-facet distribution}}.
\newblock \bibinfo{journal}{\emph{{Computer Graphics Forum}}}
  \bibinfo{volume}{31}, \bibinfo{number}{4} (\bibinfo{year}{2012}),
  \bibinfo{pages}{1509--1518}.
\newblock


\bibitem[\protect\citeauthoryear{Beckmann and Spizzichino}{Beckmann and
  Spizzichino}{1963}]%
        {BeckmannSpizzichino:1963}
\bibfield{author}{\bibinfo{person}{P. Beckmann} {and} \bibinfo{person}{A.
  Spizzichino}.} \bibinfo{year}{1963}\natexlab{}.
\newblock \bibinfo{booktitle}{\emph{The scattering of electromagnetic waves
  from rough surfaces}}.
\newblock \bibinfo{publisher}{Pergamon Press.}
\newblock


\bibitem[\protect\citeauthoryear{Cook and Torrance}{Cook and Torrance}{1982}]%
        {Cook1982REFLEC}
\bibfield{author}{\bibinfo{person}{Robert~L. Cook} {and}
  \bibinfo{person}{Kenneth~E. Torrance}.} \bibinfo{year}{1982}\natexlab{}.
\newblock \showarticletitle{A Reflectance Model for Computer Graphics}.
\newblock \bibinfo{journal}{\emph{ACM Trans. Graph.}} \bibinfo{volume}{1},
  \bibinfo{number}{1} (\bibinfo{date}{Jan.} \bibinfo{year}{1982}),
  \bibinfo{pages}{7–24}.
\newblock


\bibitem[\protect\citeauthoryear{Dupuy, Heitz, and d'Eon}{Dupuy
  et~al\mbox{.}}{2016}]%
        {Dupuy:2016:Unification}
\bibfield{author}{\bibinfo{person}{Jonathan Dupuy}, \bibinfo{person}{Eric
  Heitz}, {and} \bibinfo{person}{Eugene d'Eon}.}
  \bibinfo{year}{2016}\natexlab{}.
\newblock \showarticletitle{{Additional Progress Towards the Unification of
  Microfacet and Microflake Theories}}. In
  \bibinfo{booktitle}{\emph{Eurographics Symposium on Rendering - Experimental
  Ideas \& Implementations}}. \bibinfo{publisher}{The Eurographics
  Association}, \bibinfo{pages}{55--63}.
\newblock


\bibitem[\protect\citeauthoryear{Dupuy, Heitz, Iehl, Poulin, and
  Ostromoukhov}{Dupuy et~al\mbox{.}}{2015}]%
        {dupuy:2015}
\bibfield{author}{\bibinfo{person}{Jonathan Dupuy}, \bibinfo{person}{Eric
  Heitz}, \bibinfo{person}{Jean-Claude Iehl}, \bibinfo{person}{Pierre Poulin},
  {and} \bibinfo{person}{Victor Ostromoukhov}.}
  \bibinfo{year}{2015}\natexlab{}.
\newblock \showarticletitle{{Extracting Microfacet-based BRDF Parameters from
  Arbitrary Materials with Power Iterations}}.
\newblock \bibinfo{journal}{\emph{{Computer Graphics Forum}}}
  \bibinfo{volume}{34}, \bibinfo{number}{4} (\bibinfo{year}{2015}),
  \bibinfo{pages}{21--30}.
\newblock


\bibitem[\protect\citeauthoryear{Falster, Jarabo, and Frisvad}{Falster
  et~al\mbox{.}}{2020}]%
        {Falster:2020:Wave}
\bibfield{author}{\bibinfo{person}{V. Falster}, \bibinfo{person}{A. Jarabo},
  {and} \bibinfo{person}{J.~R. Frisvad}.} \bibinfo{year}{2020}\natexlab{}.
\newblock \showarticletitle{Computing the Bidirectional Scattering of a
  Microstructure Using Scalar Diffraction Theory and Path Tracing}.
\newblock \bibinfo{journal}{\emph{Computer Graphics Forum}}
  \bibinfo{volume}{39}, \bibinfo{number}{7} (\bibinfo{year}{2020}),
  \bibinfo{pages}{231--242}.
\newblock


\bibitem[\protect\citeauthoryear{Gamboa, Gruson, and Nowrouzezahrai}{Gamboa
  et~al\mbox{.}}{2020}]%
        {Gamboa:2020:EfficientLayered}
\bibfield{author}{\bibinfo{person}{Luis~E. Gamboa}, \bibinfo{person}{Adrien
  Gruson}, {and} \bibinfo{person}{Derek Nowrouzezahrai}.}
  \bibinfo{year}{2020}\natexlab{}.
\newblock \showarticletitle{{An Efficient Transport Estimator for Complex
  Layered Materials}}.
\newblock \bibinfo{journal}{\emph{Computer Graphics Forum}}
  \bibinfo{volume}{39}, \bibinfo{number}{2} (\bibinfo{year}{2020}),
  \bibinfo{pages}{363--371}.
\newblock


\bibitem[\protect\citeauthoryear{Guo, Ha\v{s}an, and Zhao}{Guo
  et~al\mbox{.}}{2018}]%
        {Guo:2018:Layered}
\bibfield{author}{\bibinfo{person}{Yu Guo}, \bibinfo{person}{Milo\v{s}
  Ha\v{s}an}, {and} \bibinfo{person}{Shuang Zhao}.}
  \bibinfo{year}{2018}\natexlab{}.
\newblock \showarticletitle{Position-Free Monte Carlo Simulation for Arbitrary
  Layered BSDFs}.
\newblock \bibinfo{journal}{\emph{ACM Trans. Graph.}} \bibinfo{volume}{37},
  \bibinfo{number}{6}, Article \bibinfo{articleno}{279} (\bibinfo{date}{Dec.}
  \bibinfo{year}{2018}), \bibinfo{numpages}{14}~pages.
\newblock


\bibitem[\protect\citeauthoryear{Heitz}{Heitz}{2014}]%
        {HEITZ2014UNDER}
\bibfield{author}{\bibinfo{person}{Eric Heitz}.}
  \bibinfo{year}{2014}\natexlab{}.
\newblock \showarticletitle{Understanding the Masking-Shadowing Function in
  Microfacet-Based BRDFs}.
\newblock \bibinfo{journal}{\emph{Journal of Computer Graphics Techniques
  (JCGT)}} \bibinfo{volume}{3}, \bibinfo{number}{2} (\bibinfo{date}{June}
  \bibinfo{year}{2014}), \bibinfo{pages}{48--107}.
\newblock


\bibitem[\protect\citeauthoryear{Heitz and d'Eon}{Heitz and d'Eon}{2014}]%
        {Heitz2014imporSAM}
\bibfield{author}{\bibinfo{person}{Eric Heitz} {and} \bibinfo{person}{Eugene
  d'Eon}.} \bibinfo{year}{2014}\natexlab{}.
\newblock \showarticletitle{Importance Sampling Microfacet-Based BSDFs using
  the Distribution of Visible Normals}.
\newblock \bibinfo{journal}{\emph{Computer Graphics Forum}}
  \bibinfo{volume}{33}, \bibinfo{number}{4} (\bibinfo{year}{2014}),
  \bibinfo{pages}{103--112}.
\newblock


\bibitem[\protect\citeauthoryear{Heitz and Dupuy}{Heitz and Dupuy}{2015}]%
        {heitz2015implement}
\bibfield{author}{\bibinfo{person}{Eric Heitz} {and} \bibinfo{person}{Jonathan
  Dupuy}.} \bibinfo{year}{2015}\natexlab{}.
\newblock \bibinfo{title}{Implementing a Simple Anisotropic Rough Diffuse
  Material with Stochastic Evaluation}.
\newblock
  \bibinfo{howpublished}{\url{https://drive.google.com/file/d/0BzvWIdpUpRx_M3ZmakxHYXZWaUk/view}}.
\newblock


\bibitem[\protect\citeauthoryear{Heitz, Hanika, d'Eon, and Dachsbacher}{Heitz
  et~al\mbox{.}}{2016}]%
        {heitz2016}
\bibfield{author}{\bibinfo{person}{Eric Heitz}, \bibinfo{person}{Johannes
  Hanika}, \bibinfo{person}{Eugene d'Eon}, {and} \bibinfo{person}{Carsten
  Dachsbacher}.} \bibinfo{year}{2016}\natexlab{}.
\newblock \showarticletitle{Multiple-Scattering Microfacet BSDFs with the Smith
  Model}.
\newblock \bibinfo{journal}{\emph{ACM Trans. Graph.}} \bibinfo{volume}{35},
  \bibinfo{number}{4}, Article \bibinfo{articleno}{58} (\bibinfo{date}{July}
  \bibinfo{year}{2016}), \bibinfo{numpages}{14}~pages.
\newblock


\bibitem[\protect\citeauthoryear{Jakob}{Jakob}{2010}]%
        {Mitsuba}
\bibfield{author}{\bibinfo{person}{Wenzel Jakob}.}
  \bibinfo{year}{2010}\natexlab{}.
\newblock \bibinfo{title}{Mitsuba renderer}.
\newblock
\newblock
\newblock
\shownote{http://www.mitsuba-renderer.org.}


\bibitem[\protect\citeauthoryear{Jakob, d'Eon, Jakob, and Marschner}{Jakob
  et~al\mbox{.}}{2014}]%
        {jakob:2014:layered}
\bibfield{author}{\bibinfo{person}{Wenzel Jakob}, \bibinfo{person}{Eugene
  d'Eon}, \bibinfo{person}{Otto Jakob}, {and} \bibinfo{person}{Steve
  Marschner}.} \bibinfo{year}{2014}\natexlab{}.
\newblock \showarticletitle{A Comprehensive Framework for Rendering Layered
  Materials}.
\newblock \bibinfo{journal}{\emph{ACM Trans. Graph.}} \bibinfo{volume}{33},
  \bibinfo{number}{4}, Article \bibinfo{articleno}{118} (\bibinfo{date}{July}
  \bibinfo{year}{2014}), \bibinfo{numpages}{14}~pages.
\newblock


\bibitem[\protect\citeauthoryear{Kelemen and Szirmay-Kalos}{Kelemen and
  Szirmay-Kalos}{2001}]%
        {Kelemen01amicrofacet}
\bibfield{author}{\bibinfo{person}{Csaba Kelemen} {and}
  \bibinfo{person}{László Szirmay-Kalos}.} \bibinfo{year}{2001}\natexlab{}.
\newblock \showarticletitle{A Microfacet Based Coupled Specular-Matte BRDF
  Model with Importance Sampling}. In \bibinfo{booktitle}{\emph{Eurographics
  2001 -- Short Presentations}}. \bibinfo{publisher}{The Eurographics
  Association}.
\newblock


\bibitem[\protect\citeauthoryear{Kulla and Conty}{Kulla and Conty}{2017}]%
        {KullaConty:2017:revisiting}
\bibfield{author}{\bibinfo{person}{Christopher Kulla} {and}
  \bibinfo{person}{Alejandro Conty}.} \bibinfo{year}{2017}\natexlab{}.
\newblock \bibinfo{title}{Physically Based Shading in Theory and Practice -
  Revisiting Physically Based Shading at Imageworks}.
\newblock
\newblock
\newblock
\shownote{http://blog.selfshadow.com/publications/s2017-shading-course/.}


\bibitem[\protect\citeauthoryear{Lee, Jarabo, Jeon, Gutiérrez, and Kim}{Lee
  et~al\mbox{.}}{2018}]%
        {lEE2018PRATMUL}
\bibfield{author}{\bibinfo{person}{Joo Lee}, \bibinfo{person}{Adrián Jarabo},
  \bibinfo{person}{Daniel Jeon}, \bibinfo{person}{Diego Gutiérrez}, {and}
  \bibinfo{person}{Min Kim}.} \bibinfo{year}{2018}\natexlab{}.
\newblock \showarticletitle{Practical multiple scattering for rough surfaces}.
\newblock \bibinfo{journal}{\emph{ACM Trans. Graph.}}  \bibinfo{volume}{37},
  Article \bibinfo{articleno}{175} (\bibinfo{date}{Dec.} \bibinfo{year}{2018}),
  \bibinfo{numpages}{12}~pages.
\newblock


\bibitem[\protect\citeauthoryear{{Meneveaux}, {Bringier}, {Tauzia},
  {Ribardière}, and {Simonot}}{{Meneveaux} et~al\mbox{.}}{2018}]%
        {Meneveaux:2018}
\bibfield{author}{\bibinfo{person}{Daniel {Meneveaux}},
  \bibinfo{person}{Benjamin {Bringier}}, \bibinfo{person}{Emmanuelle {Tauzia}},
  \bibinfo{person}{Mickaël {Ribardière}}, {and} \bibinfo{person}{Lionel
  {Simonot}}.} \bibinfo{year}{2018}\natexlab{}.
\newblock \showarticletitle{Rendering Rough Opaque Materials with Interfaced
  Lambertian Microfacets}.
\newblock \bibinfo{journal}{\emph{IEEE Transactions on Visualization and
  Computer Graphics}} \bibinfo{volume}{24}, \bibinfo{number}{3}
  (\bibinfo{year}{2018}), \bibinfo{pages}{1368--1380}.
\newblock


\bibitem[\protect\citeauthoryear{Ribardi\`{e}re, Bringier, Meneveaux, and
  Simonot}{Ribardi\`{e}re et~al\mbox{.}}{2017}]%
        {Ribardiere:2017:STD}
\bibfield{author}{\bibinfo{person}{Micka\"{e}l Ribardi\`{e}re},
  \bibinfo{person}{Benjamin Bringier}, \bibinfo{person}{Daniel Meneveaux},
  {and} \bibinfo{person}{Lionel Simonot}.} \bibinfo{year}{2017}\natexlab{}.
\newblock \showarticletitle{STD: Student's t-Distribution of Slopes for
  Microfacet Based BSDFs}.
\newblock \bibinfo{journal}{\emph{Computer Graphics Forum}}
  \bibinfo{volume}{36}, \bibinfo{number}{2} (\bibinfo{year}{2017}),
  \bibinfo{pages}{421--429}.
\newblock


\bibitem[\protect\citeauthoryear{Ribardi\`{e}re, Bringier, Simonot, and
  Meneveaux}{Ribardi\`{e}re et~al\mbox{.}}{2019}]%
        {Ribardiere:2019:NDF}
\bibfield{author}{\bibinfo{person}{Micka\"{e}l Ribardi\`{e}re},
  \bibinfo{person}{Benjamin Bringier}, \bibinfo{person}{Lionel Simonot}, {and}
  \bibinfo{person}{Daniel Meneveaux}.} \bibinfo{year}{2019}\natexlab{}.
\newblock \showarticletitle{Microfacet BSDFs Generated from NDFs and Explicit
  Microgeometry}.
\newblock \bibinfo{journal}{\emph{ACM Trans. Graph.}} \bibinfo{volume}{38},
  \bibinfo{number}{5}, Article \bibinfo{articleno}{143} (\bibinfo{date}{June}
  \bibinfo{year}{2019}), \bibinfo{numpages}{15}~pages.
\newblock


\bibitem[\protect\citeauthoryear{Sch\"{u}ssler, Heitz, Hanika, and
  Dachsbacher}{Sch\"{u}ssler et~al\mbox{.}}{2017}]%
        {Schussler:2017:normal}
\bibfield{author}{\bibinfo{person}{Vincent Sch\"{u}ssler},
  \bibinfo{person}{Eric Heitz}, \bibinfo{person}{Johannes Hanika}, {and}
  \bibinfo{person}{Carsten Dachsbacher}.} \bibinfo{year}{2017}\natexlab{}.
\newblock \showarticletitle{Microfacet-Based Normal Mapping for Robust Monte
  Carlo Path Tracing}.
\newblock \bibinfo{journal}{\emph{ACM Trans. Graph.}} \bibinfo{volume}{36},
  \bibinfo{number}{6}, Article \bibinfo{articleno}{205} (\bibinfo{year}{2017}),
  \bibinfo{numpages}{12}~pages.
\newblock


\bibitem[\protect\citeauthoryear{{Smith}}{{Smith}}{1967}]%
        {smith:1967:smith}
\bibfield{author}{\bibinfo{person}{B. {Smith}}.}
  \bibinfo{year}{1967}\natexlab{}.
\newblock \showarticletitle{Geometrical shadowing of a random rough surface}.
\newblock \bibinfo{journal}{\emph{IEEE Transactions on Antennas and
  Propagation}} \bibinfo{volume}{15}, \bibinfo{number}{5}
  (\bibinfo{year}{1967}), \bibinfo{pages}{668--671}.
\newblock


\bibitem[\protect\citeauthoryear{Torrance and Sparrow}{Torrance and
  Sparrow}{1967}]%
        {TorranceTorrance:1967}
\bibfield{author}{\bibinfo{person}{Kenneth Torrance} {and} \bibinfo{person}{E.
  Sparrow}.} \bibinfo{year}{1967}\natexlab{}.
\newblock \showarticletitle{Theory for Off-Specular Reflection From Roughened
  Surfaces}.
\newblock \bibinfo{journal}{\emph{Journal of The Optical Society of America}}
  \bibinfo{volume}{57} (\bibinfo{date}{Sept.} \bibinfo{year}{1967}).
\newblock


\bibitem[\protect\citeauthoryear{Trowbridge and Reitz}{Trowbridge and
  Reitz}{1975}]%
        {Trowbridge:75}
\bibfield{author}{\bibinfo{person}{T.~S. Trowbridge} {and}
  \bibinfo{person}{K.~P. Reitz}.} \bibinfo{year}{1975}\natexlab{}.
\newblock \showarticletitle{Average irregularity representation of a rough
  surface for ray reflection}.
\newblock \bibinfo{journal}{\emph{Journal of The Optical Society of America}}
  \bibinfo{volume}{65}, \bibinfo{number}{5} (\bibinfo{date}{May}
  \bibinfo{year}{1975}), \bibinfo{pages}{531--536}.
\newblock


\bibitem[\protect\citeauthoryear{Turquin}{Turquin}{2019}]%
        {Turquin:2019:multiple}
\bibfield{author}{\bibinfo{person}{Emmanuel Turquin}.}
  \bibinfo{year}{2019}\natexlab{}.
\newblock \bibinfo{title}{Practical multiple scattering compensation for
  microfacet models}.
\newblock
  \bibinfo{howpublished}{\url{https://blog.selfshadow.com/publications/turquin/ms_comp_final.pdf}}.
\newblock


\bibitem[\protect\citeauthoryear{Walter, Marschner, Li, and Torrance}{Walter
  et~al\mbox{.}}{2007}]%
        {walter2007mmrt}
\bibfield{author}{\bibinfo{person}{Bruce Walter}, \bibinfo{person}{Stephen~R.
  Marschner}, \bibinfo{person}{Hongsong Li}, {and} \bibinfo{person}{Kenneth~E.
  Torrance}.} \bibinfo{year}{2007}\natexlab{}.
\newblock \showarticletitle{{Microfacet Models for Refraction through Rough
  Surfaces}}. In \bibinfo{booktitle}{\emph{Rendering Techniques (proc. EGSR
  2007)}}. \bibinfo{publisher}{The Eurographics Association},
  \bibinfo{pages}{195--206}.
\newblock


\bibitem[\protect\citeauthoryear{Westin, Arvo, and Torrance}{Westin
  et~al\mbox{.}}{1992}]%
        {Westin:92:RandomWalk}
\bibfield{author}{\bibinfo{person}{Stephen~H. Westin},
  \bibinfo{person}{James~R. Arvo}, {and} \bibinfo{person}{Kenneth~E.
  Torrance}.} \bibinfo{year}{1992}\natexlab{}.
\newblock \showarticletitle{Predicting Reflectance Functions from Complex
  Surfaces}. In \bibinfo{booktitle}{\emph{Proceedings of the 19th Annual
  Conference on Computer Graphics and Interactive Techniques}}
  \emph{(\bibinfo{series}{SIGGRAPH '92})}. \bibinfo{publisher}{Association for
  Computing Machinery}, \bibinfo{pages}{255–264}.
\newblock


\bibitem[\protect\citeauthoryear{Xia, Walter, Hery, and Marschner}{Xia
  et~al\mbox{.}}{2020}]%
        {Xia:2020:Layered}
\bibfield{author}{\bibinfo{person}{Mengqi~(Mandy) Xia}, \bibinfo{person}{Bruce
  Walter}, \bibinfo{person}{Christophe Hery}, {and} \bibinfo{person}{Steve
  Marschner}.} \bibinfo{year}{2020}\natexlab{}.
\newblock \showarticletitle{Gaussian Product Sampling for Rendering Layered
  Materials}.
\newblock \bibinfo{journal}{\emph{Computer Graphics Forum}}
  \bibinfo{volume}{39}, \bibinfo{number}{1} (\bibinfo{year}{2020}),
  \bibinfo{pages}{420--435}.
\newblock


\bibitem[\protect\citeauthoryear{Xie and Hanrahan}{Xie and Hanrahan}{2018}]%
        {Feng2018multiV}
\bibfield{author}{\bibinfo{person}{Feng Xie} {and} \bibinfo{person}{Pat
  Hanrahan}.} \bibinfo{year}{2018}\natexlab{}.
\newblock \showarticletitle{Multiple Scattering from Distributions of Specular
  V-Grooves}.
\newblock \bibinfo{journal}{\emph{ACM Trans. Graph.}} \bibinfo{volume}{37},
  \bibinfo{number}{6}, Article \bibinfo{articleno}{276} (\bibinfo{year}{2018}),
  \bibinfo{numpages}{14}~pages.
\newblock


\bibitem[\protect\citeauthoryear{Xie, Kaplanyan, Hunt, and Hanrahan}{Xie
  et~al\mbox{.}}{2019}]%
        {Xie:2019:multiple}
\bibfield{author}{\bibinfo{person}{Feng Xie}, \bibinfo{person}{Anton
  Kaplanyan}, \bibinfo{person}{Warren Hunt}, {and} \bibinfo{person}{Pat
  Hanrahan}.} \bibinfo{year}{2019}\natexlab{}.
\newblock \showarticletitle{Multiple Scattering Using Machine Learning}. In
  \bibinfo{booktitle}{\emph{ACM SIGGRAPH 2019 Talks}}. Article
  \bibinfo{articleno}{70}, \bibinfo{numpages}{2}~pages.
\newblock


\end{thebibliography}
% Appendix
\appendix
\end{document}